\documentclass[a4paper,11pt]{article}

\usepackage[a4paper]{geometry}

\usepackage{amsmath,amsfonts,amssymb}
\usepackage{mathtools}
\usepackage{graphicx}
\usepackage[font=small,labelfont=bf]{caption}
\usepackage[linktoc=all]{hyperref}
\usepackage[numbers,sort&compress]{natbib}
\usepackage{doi}
\usepackage{url}

\begin{document}

\title{Efficient Reconciliation of Continuous Variable Quantum Key Distribution with Multiplicatively Repeated Non-Binary LDPC Codes}

\author{Jesus Martinez-Mateo and David Elkouss
\thanks{Jesus Martinez-Mateo is with the Departamento de Matem\'{a}tica Aplicada a las Tecnolog\'{i}as de la Informaci\'{o}n y las Comunicaciones, Universidad Polit\'{e}cnica de Madrid, Spain (e-mail: jesus.martinez.mateo@upm.es). David Elkouss is with the Networked Quantum Devices Unit, Okinawa Institute of Science and Technology Graduate University, Japan (e-mail: david.elkouss@oist.jp).}%
}

\date{}

\maketitle

\begin{abstract}
Continuous variable quantum key distribution bears the promise of simple quantum key distribution directly compatible with commercial off the shelf equipment. However, for a long time its performance was hindered by the absence of good classical postprocessing capable of distilling secret-keys in the noisy regime. Advanced coding solutions in the past years have partially addressed this problem enabling record transmission distances of up to 165~km, and 206~km over ultra-low loss fiber. In this paper, we show that a very simple coding solution with a single code is sufficient to extract keys at all noise levels. This solution has performance competitive with prior results for all levels of noise, and we show that non-zero keys can be distilled up to a record distance of 192~km assuming the standard loss of a single-mode optical fiber, and 240~km over ultra-low loss fibers. Low-rate codes are constructed using multiplicatively repeated non-binary low-density parity-check codes over a finite field of characteristic two. This construction only makes use of a $(2, k)$-regular non-binary low-density parity-check code as mother code, such that code design is in fact not required, thus trivializing the code construction procedure. The construction is also inherently rate-adaptive thereby allowing to easily create codes of any rate. Rate-adaptive codes are of special interest for the efficient reconciliation of errors over time or arbitrary varying channels, as is the case with quantum key distribution. In short, these codes are highly efficient when reconciling errors over a very noisy communication channel, and perform well even for short block-length codes. Finally, the proposed solution is known to be easily amenable to hardware implementations, thus addressing also the requirements for practical reconciliation in continuous variable quantum key distribution.
\end{abstract}

%\keywords{Continuous variable quantum key distribution, Information reconciliation, Low-rate coding, Non-binary low-density parity-check codes.}

\section{Introduction}

Quantum key distribution (QKD) \cite{Gisin_02} allows two distant parties, typically named Alice and Bob, communicating through a quantum channel to exchange an information-theoretically secure key. As usual in quantum communications, quantum states carrying information are encoded into photons and transmitted via fiber optics between the parties. However, due to imperfections when preparing and measuring the transmitted quantum states, and also due to noise in the quantum channel, there may be disparities (errors) in the exchanged raw keys---that must be assumed to be caused by any hypothetical eavesdropper. In consequence, once the raw key exchange or key generation process is concluded, a key distillation process has to be performed to convert their correlated but noisy raw keys into a shared, error free, secret key. Then, some information from the raw keys needs first to be disclosed during an information reconciliation (error correction) procedure \cite{Brassard_94}, carried out over a public noiseless and authenticated channel, to produce a common string, that is, an identical key on both sides. Subsequently, some information needs to be removed in a privacy amplification procedure \cite{Bennett_88} to produce a shorter, but secret, key. Therefore, to maximize the secret key rate and to achieve greater distances between the parties, highly efficient information reconciliation methods are necessary in every experimental realization of QKD.

Discrete-variable (DV) QKD makes use of discrete modulation of quantum states, and generates correlated discrete variables at Alice's and Bob's sides. In a typical DV-QKD protocol, such as the well-known BB84 proposed by Bennett and Brassard in 1984 \cite{Bennett_84}, each quantum state encodes a single bit, so that the exchanged raw keys are correlated bit strings. In such a context, standard binary linear codes, that have been demonstrated to be highly efficient, can be used for reconciling errors in the exchanged keys. Several examples of efficient information reconciliation methods have been proposed using, for instance, low-density parity-check codes \cite{Elkouss_11, Martinez_12, Martinez_13, Tarable_24} and polar codes \cite{Jouguet_14}. Other reconciliation methods, such as Cascade and its modified versions \cite{Martinez_15, Pacher_15}, have also been demonstrated to be highly efficient despite not using conventional decoding techniques.

In continuous-variable (CV) QKD the situation is significantly different. CV-QKD makes use of continuous modulation of quantum states, and generates correlated Gaussian variables at Alice's and Bob's sides. Contrary to what happens in DV-QKD, these devices typically operate in the regime of low signal-to-noise ratio (SNR), that is, the information is transmitted over a very noisy communication channel. Moreover, the reconciliation efficiency of correlated Gaussian variables is decisive when determining the achievable secret key rate, thus limiting the maximum attainable distance between the parties. Unfortunately, correcting errors with low-rate codes is relatively complicated, since the efficiency of common decoding techniques drops in the low SNR regime and also the decoding complexity is significantly increased. For CV-QKD with Gaussian modulation several approaches have been explored to improve the reconciliation efficiency of correlated Gaussian variables. Originally, a slice reconciliation scheme was proposed in~\cite{VanAssche_04}. This scheme divides a continuous function into slices that are reconciled independently using codes of different rates, being thus compatible with existing standard binary low-density parity-check and polar codes. Later, another approach called multidimensional reconciliation was proposed in~\cite{Leverrier_08}, with the idea of reducing the problem of reconciling correlated Gaussian variables to the well-known channel coding problem over the additive white Gaussian noise (AWGN) channel. In this approach, the physical Gaussian channel is transformed into a channel close to the binary-input AWGN channel. Numerous alternatives have been proposed for information reconciliation in CV-QKD, most of them based on low-density parity-check codes (considering both schemes, slice and multidimensional reconciliation). Multi-edge type low-density parity-check codes are probably the most widely used method for reconciling errors in the low SNR regime \cite{Jouguet_11, Jouguet_14b, Bai_17, Wang_18, ZhangYichen_20, Mani_21}, sometimes also considering rate-adaptive techniques for highly efficient decoding \cite{Jiang_17, Wang_17, Jeong_22}, or quasi-cyclic codes for layered and fast decoding \cite{Milicevic_18}. Additionally, it can also be found proposals that use other codes, such as repeat-accumulate codes \cite{Johnson_16}, raptor codes \cite{Shirvanimoghaddam_16, Zhou_19}, or polar codes \cite{Jouguet_14, Zhang_24} among others.

Traditional DV-QKD and CV-QKD protocols were severely limited in distance because the transmittance of the quantum channel decreases exponentially with distance. However, recent advances, both theoretical and experimental, have made it possible to extend this distance to a few hundred kilometers. A recent discrete-variable proposal called twin-field QKD (TF-QKD) allows one to extend this distance by changing the fundamental scaling of the rate \cite{lucamarini2018overcoming} with state-of-the-art demonstrations reaching up to approximately one thousand kilometers \cite{Chen_20, Wang_22, liu2023experimental}.

In this contribution, we discuss a family of low-rate codes that efficiently correct errors on the binary-input AWGN channel, even at low and ultra low SNR regime. These codes are constructed from $(2,k)$-regular non-binary low-density parity-check codes, and their construction is quite simple since no code design is required. These low-rate codes are of particular interest for CV-QKD given that their decoding is highly efficient and their construction is also inherently rate-adaptive, that is, they remain efficient even when the channel varies or the channel noise is different. Furthermore, the proposed codes and their decoding are suitable for hardware implementations as demonstrated in~\cite{Ferraz_22}.

The remainder of this paper is organized as follows. In Section~\ref{sec:background} we review the background of binary and non-binary low-density parity-check codes and their interest for correcting or reconciling errors in QKD. Then, we describe the proposed low-rate code construction and the corresponding efficient decoding algorithm. We validate the construction with comprehensive numerical simulations in Section~\ref{sec:results}. Finally, we present our conclusions in Section~\ref{sec:conclusions}.

\section{Background} 
\label{sec:background}

\subsection{Information Reconciliation with Non-Binary Low-Density Parity-Check Codes} 
\label{sec:nonbinary-ldpc}

Low-density parity-check (LDPC) codes were introduced by Gallager in the early 1960s \cite{Gallager_63}, but remained largely unexplored until the late 90s, when MacKay and Neal revisited these codes and explored their potential \cite{MacKay_96, MacKay_99}. It soon turned out that their performance over binary input memoryless channels was very close to channel capacity \cite{Richardson_01b}, that is, they have good thresholds. Furthermore, the sparsity of LDPC matrices allows high performance and low complexity decoding using iterative message-passing algorithms based on belief propagation (such as the sum-product algorithm). Other decoding algorithms and techniques (for instance, normalized and offset min-sum algorithms, or serial and parallel schedule) also exhibit a good trade-off between performance and complexity, making them suitable for hardware implementations. Consequently, LDPC codes were regarded as one of the most promising coding techniques.

Non-binary LDPC codes were also considered by Gallager in his seminal work \cite{Gallager_63}. Davey and MacKay later found that these codes can outperform binary LDPC codes \cite{Davey_98}. However, this improvement was achieved at the expense of increased decoding complexity. Several low-complexity algorithms were then proposed \cite{Barnault_03, Voicila_10, Ferraz_22} for decoding non-binary LDPC codes over finite (Galois) fields of order $q$, in the following denoted by $\operatorname{GF}(q)$, where $q$ is a power of prime $q = p^m$, $q>2$ with $p$ prime and positive integer $m \geq 1$. Thereafter, these codes have attracted much attention thanks to their good performance.

As linear codes, binary and non-binary LDPC codes can be represented by a Tanner or bipartite graph, that is, a graphical depiction of a parity-check matrix $H$. Nodes in a Tanner graph are divided between symbol (or variable) nodes and check (or constraint) nodes. Each check node corresponds to a parity-check constraint (that is, an implicit equation of a linear code, a row in $H$), and thus it is connected by edges to those variables (symbols nodes) involved in the equation. This representation is useful for both decoding and code construction. On the construction of these codes, it is well known that good binary LDPC codes are designed allowing symbols to participate in different checks in an irregular fashion \cite{Richardson_01a}. Therefore, each symbol node connects to a number of check nodes, and we refer to this as symbol node degree. Such codes are referred to as irregular LDPC codes, and good code ensembles are designed by optimizing symbol and check node degree distributions on Tanner graphs \cite{Richardson_01b, Chung_01}. However, while irregular Tanner graphs help improve the performance of binary LDPC codes, this is not the case for non-binary LDPC codes, where $(2,k)$-regular non-binary LDPC codes over $\operatorname{GF}(q)$ are empirically known to be the best performing codes \cite{Polliat_08}. This is another significant advantage of non-binary LDPC codes.

Based on both binary and non-binary LDPC codes efficient methods have been proposed for information reconciliation in DV-QKD, all of them in the high-rate regime (that is, for code rates greater than one half). Notable examples are, for instance, highly efficient reconciliation methods using rate-adaptive binary \cite{Elkouss_11} and non-binary LDPC codes \cite{Kasai_10b, Mueller_24}, blind (or interactive) reconciliation using also rate-adaptive but short block-length LDPC codes \cite{Martinez_12, Kiktenko_17, Liu_20}, and high-throughput reconciliation with layered decoding and quasi-cyclic LDPC codes \cite{Martinez_13}. These methods can also be used for information reconciliation in CV-QKD with slice reconciliation.

Binary and non-binary LDPC codes have been empirically shown to have good performance for high-rate codes. However, when transmitting information over a very noisy communication channel, that is, for low-rate coding, their performance degrades rapidly \cite{Andriyanova_12}. Multi-edge type LDPC codes were originally proposed by Richardson and Urbanke to design high-rate codes with low error floors and low-rate codes with high performance \cite{Richardson_04}. These codes are an extension of standard binary LDPC codes where in the Tanner graph only one type of edges is considered. In multi-edge type LDPC codes, there are different edge types such that a node is no longer characterized by a single degree but by a vector degree. Unfortunately, a major shortcoming of low-rate multi-edge type LDPC codes is the large number of check node computations which significantly increases their decoding complexity.

Efficient reconciliation methods based on multi-edge type LDPC codes have been proposed for information reconciliation in CV-QKD \cite{Jouguet_11, Jouguet_14b, Johnson_17, Bai_17, Jiang_17, Wang_17, Wang_18, Milicevic_18, Mani_21}. However, most of these proposals consider very large block-length codes of approximately $10^6$ bits, or even larger, which make hardware
implementations unrealistic. Therefore, these approaches neither allow for efficient decoding using short or intermediate block-length codes, nor can be efficiently implemented. Their decoding complexity can only be reduced by limiting the maximum check node degree that may lead to suboptimal codes \cite{Suhwang_19}.

However, we consider that the potential of non-binary LDPC codes has not been explored enough in the context of CV-QKD, that is, for reconciling continuous correlated variables\footnote{Except for the case of the correlated bivariate normal distribution \cite{Pacher_16}.}. The purpose of this work is to study the possibility of adapting non-binary LDPC codes to efficiently operate in the low SNR regime. To this end, we carried out a comprehensive analysis of the multiplicatively repeated non-binary LDPC codes proposed in~\cite{Kasai_11}, and in this paper we show their interest for low-rate coding and their application in CV-QKD postprocessing. In the following, we first describe these codes in Section~\ref{sec:multirep}, and then a modified decoding algorithm for reconciling errors is given in Section~\ref{sec:decoding}.

\subsection{Multiplicatively Repeated Non-Binary LDPC Codes}
\label{sec:multirep}

Kasai and Declerq proposed in~\cite{Kasai_11} the concatenation of non-binary LDPC codes with multiplicative repetition inner codes. According to their proposal, a $(2,k)$-regular non-binary LDPC code over a finite field of order $2^p$ is multiplicatively repeated to construct non-binary LDPC codes of lower rates. Surprisingly, as we show below, such simple low-rate non-binary LDPC code construction using high order fields outperforms other low-rate codes so far, particularly when considering short and intermediate block-length codes.

\begin{figure}
\centering
\includegraphics[width=0.35\linewidth]{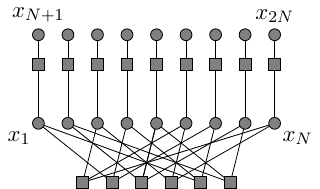}
\caption{Example of $C_2$. A $(2,3)$-regular linear (mother) code is concatenated with an inner multiplicative repetition code of length $2$, resulting in a code of rate $1/6$.}
\label{fig:example-c2}
\end{figure}

Let $C_1$ be a $(2,3)$-regular non-binary LDPC code over $\operatorname{GF}(2^p)$ of length $N$ symbols, or equivalently $Np$ bits\footnote{For convenience and faster decoding \cite{Barnault_03}, we only consider finite fields of characteristic two, that is, binary finite fields of order $2^p$. The elements of the finite field are binary polynomials of degree less than or equal to $p-1$, that is, polynomials whose coefficients are either $0$ or $1$. Operations in the finite field (that is, addition and multiplication, and their inverse operations, subtraction and division, respectively) can then be efficiently implemented.}, and rate $1/3$. In the following, we refer to this code as \emph{mother code}. From this mother code we construct a code $C_2$ in the following way. We choose $N$ coefficients $r_{N+1}, \ldots, r_{2N}$ uniformly at random from the finite set $\operatorname{GF}(2^p) \setminus \{0\}$, then for each codeword in $C_1$ we define a codeword in $C_2$ as follows:
\[
\begin{split}
C_2 = \left\{ (x_1, x_2, \ldots, x_{2N}) : (x_1, \ldots, x_N) \in C_1, \right. \\
\left. x_{N+n} = r_{N+n} x_n \;\text{for}\; n=1,\ldots,N \right\}.
\end{split}
\]

We say that the symbol $x_{N+n}$ (with $x_{N+n} = r_{N+n} x_n$) is a \emph{multiplicative repetition} symbol of $x_n$, for $n = 1, \ldots, N$. Therefore, we construct $C_2$ by multiplicatively repeating each symbol node of the mother code $C_1$. Note also that, for each multiplicative repetition symbol we have an additional parity-check constraint, that is, it holds $x_{N+n} + r_{N+n} x_n = 0$ for $n = 1, \ldots, N$. Hence, in other words, we construct the code $C_2$ by connecting each symbol node $x_n$ of $C_1$ to a new check node (the additional parity-check constraint), which is also connected to a second symbol node $x_{N+n}$ (the multiplicative repetition symbol of $x_n$). Figure~\ref{fig:example-c2} depicts an example of $C_2$. As shown, each symbol node of degree one in the figure represents a multiplicative repetition symbol $x_{N+n}$ of $x_n$ for $n = 1, \ldots, N$, and each check node of degree two represents a new parity-check constraint.

Clearly, the constructed code $C_2$ has the same number of codewords as the mother code $C_1$, however the codewords of $C_2$ are twice as long as the codewords of $C_1$, that is, the code length of $C_2$ is $2N$ symbols, thus resulting in a lower code rate of $1/6$.

\begin{figure}
\centering
\includegraphics[width=0.55\linewidth]{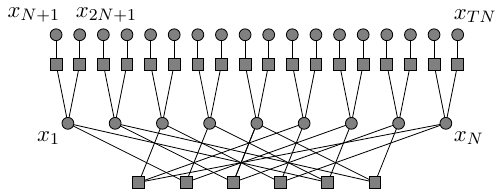}
\caption{Example of $C_3$. A $(2,3)$-regular linear code is concatenated with an inner multiplicative repetition code of length $3$. The code rate is $1/9$.}
\label{fig:example-c3}
\end{figure}

Next, from $C_2$ we construct another code $C_3$ in a similar way. We choose again $N$ coefficients $r_{2N+1}, \ldots, r_{3N}$ uniformly at random from the finite set $\operatorname{GF}(2^p) \setminus \{0\}$, and then we construct $C_3$ from $C_2$ by multiplicatively repeating each symbol of $C_1$ as follows:
\[
\begin{split}
C_3 = \left\{ (x_1, x_2, \ldots, x_{3N}) : (x_1, \ldots, x_{2N}) \in C_2, \right. \\
\left. x_{2N+n} = r_{2N+n} x_n \;\text{for}\; n=1,\ldots,N \right\}.
\end{split}
\]

$C_3$ has the same number of codewords as $C_2$ and $C_1$, but the codeword length is now of $3N$ symbols. Thus, the code rate of $C_3$ is $1/9$. Figure~\ref{fig:example-c3} depicts an example of $C_3$. Again, symbol nodes of degree one correspond to multiplicative repetition symbols, whereas check nodes of degree two correspond to parity-check constraints induced by each multiplicative repetition symbol.

Subsequent lower rate codes are constructed recursively, as we have shown above for $C_2$ and $C_3$. A code $C_T$, with $T \ge 2$, is defined recursively as follows. We choose $N$ coefficients $r_{(T-1)N+1}, \ldots, r_{TN}$ uniformly at random from $\operatorname{GF}(2^p) \setminus \{0\}$, then we construct $C_T$ from $C_{T-1}$ by multiplicatively repeating each symbol of $C_1$ as follows:
\[
\begin{split}
C_T = \left\{ (x_1, x_2, \ldots, x_{TN}) : (x_1, \ldots, x_{(T-1)N}) \in C_{T-1}, \right. \\
\left. x_{(T-1)N+n} = r_{(T-1)N+n} x_n \;\text{for}\; n=1,\ldots,N \right\}.
\end{split}
\]

The code $C_T$ has a length of $T \cdot N$ symbols, and rate $1/(3T)$. We refer to $T$ as \emph{repetition parameter}.

Note that, this code construction is inherently rate-adaptive since in the construction of the last code $C_T$ from $C_{T-1}$ we can add as many multiplicative repetition symbols as we wish, obviously between $1$ and $N$. Hence, from a code $C_{T-1}$ with codewords of length $(T-1)N$ symbols we can construct a code $C_T$ with codewords of length from $(T-1)N+1$ to $TN$ symbols, resulting in a code of rate $1/3(T-1) < R \leq 1/3T$.

\subsection{Non-Binary LDPC Decoding Algorithm}
\label{sec:decoding}

In this section we describe a belief propagation algorithm for efficiently decoding multiplicatively repeated non-binary LDPC codes\footnote{It is well-known that a belief propagation algorithm would produce the exact posterior probabilities of all the symbols after a number of iterations, that is, optimum decoding, but only if the Tanner graph contains no cycles \cite{Pearl_88}.}. The algorithm is adapted to the source coding problem with side information studied by Slepian and Wolf \cite{Slepian_73}, that more accurately describes the problem of correcting disparities between correlated sources---also known as information reconciliation or simply reconciliation in the context of secret-key agreement.

\emph{Source coding with side information:}
Let $x$ and $y$ be two correlated strings of length $N$ (that is, two strings of symbols, or equivalently, two bit strings of length $Np$) belonging to Alice and Bob, respectively, that is, these strings are realizations of the correlated sources $X$ and $Y$. The encoder computes the syndrome $z$ of his string $x$, that is, $z = Hx$ (where $H$ is the parity-check matrix of a given linear code), and sends it to the decoder through a noiseless channel. Then the decoder, given the coset index $z$ (the syndrome of $x$), look for the sequence in the coset $C_z$ that is closest to $y$, where the coset $C_z$ is a set that includes all strings of length $N$ with $z$ syndrome, that is, $C_z = \{x : Hx = z\}$. For further information see the Wyner's binning scheme \cite{Wyner_75}, or the modified decoding algorithm for binary LDPC codes proposed by Liveris \cite{Liveris_02}.

In the scenario presented above Alice is the encoder and Bob the decoder, then if Alice is also the emitter of quantum states and Bob the receiver we say that the parties perform direct reconciliation. Otherwise, that is, when the emitter and encoder are on different sides, we consider it as reverse reconciliation. To switch from one scheme to another we just need to exchange the roles (encoder and decoder).

Note that, in the following we will consider the decoding of a multiplicatively repeated non-binary LDPC code over $\operatorname{GF}(2^p)$ of length $N$ symbols. Therefore, each sequence of $p$ bits represents the binary polynomial corresponding to an element of the finite field $\operatorname{GF}(2^p)$, that is, a symbol.

\begin{figure}
\centering
\includegraphics[width=0.55\linewidth]{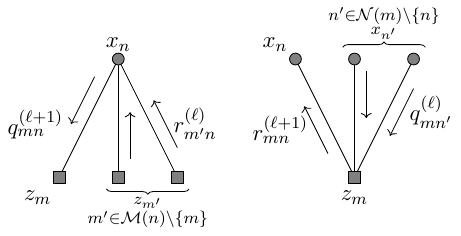}
\caption{Iterative decoding messages (probabilities) propagated along the edges of a Tanner graph. Messages $q_{mn}(\alpha)$ for all $\alpha \in \operatorname{GF}(2^p)$ are exchanged from symbol to check nodes, that is, the probability that symbol $x_n = \alpha$ given the information obtained from other checks (Left). Messages $r_{mn}(\alpha)$ for all $\alpha \in \operatorname{GF}(2^p)$ are exchanged from check to symbol nodes, that is, the probability of check $z_m$ being satisfied if symbol $x_n = \alpha$ given the information obtained from other symbols (Right). Note that, messages or probabilities in iteration $\ell+1$ are calculated from the messages exchanged in the previous iteration $\ell$.}
\label{fig:messages}
\end{figure}

\emph{Decoding algorithm:}
Let $\mathcal{N}(m)$ be the set of indexes of symbol nodes adjacent to the check node $z_m$, and let $\mathcal{M}(n)$ be the set of indexes of check nodes adjacent to the symbol node $x_n$, that is, $\mathcal{N}(m) = \{n : h_{mn} \neq 0\}$ and $\mathcal{M}(n) = \{m : h_{mn} \neq 0\}$, respectively, where $H=(h_{mn})$ is the parity-check matrix of a given linear code. An iterative decoding (belief propagation based) algorithm for non-binary LDPC codes, such as in the binary case, is a message-passing algorithm where probabilities are propagated along the edges of the Tanner graph associated with the parity-check matrix $H$. The algorithm consists mainly of two alternating steps. On each iteration $\ell$, first messages $r_{mn}^{(\ell)}$ are exchanged from check to symbols nodes, and later messages $q_{mn}^{(\ell)}$ are exchanged from symbol to check nodes. Figure~\ref{fig:messages} depicts how these messages are iteratively updated using extrinsic information, that is, probabilities obtained in a previous iteration from other neighboring nodes. Both steps are repeated until all of symbol values are known (that is, all the parity-check constraints are satisfied), or a maximum number of decoding iterations is reached.

\emph{Step 1. Initialization:}
Let $X_n$ and $Y_n$ be the random variables of the $n$-th transmitted and received symbols, respectively, and let $y_n$ be $n$-th received symbol, that is, the channel output of the $n$-th transmitted symbol. For each symbol node $x_n$ in the mother code $C_1$, with $n = 1, \ldots, N$, we calculate $2^p$ prior probabilities, that is, the \emph{a priori} probability of symbol $x_n$ being $\alpha$:
\[
p_n^{(0)}(\alpha) = \Pr(X_n=\alpha | Y_n=y_n), \quad \forall \alpha \in \operatorname{GF}(2^p).
\]

In order to calculate the a priori probability of each symbol node $x_n$ in the mother code $C_1$ we have to consider also the a priori probabilities of the multiplicative repetition symbols $x_{tN+n}$ of $x_n$, given the check constraints $x_{tN+n} = r_{tN+n} x_n$, for $t=1,\ldots,T-1$.

Then, each symbol node $x_n$ initially sends the message $q_{mn}^{(0)} = p_n^{(0)}$ to its adjacent check nodes $z_m$, for all $m \in \mathcal{M}(n)$, that is:
\[
\begin{split}
q_{mn}^{(0)}(\alpha) &= p_n^{(0)}(\alpha), \quad \forall \alpha \in \operatorname{GF}(2^p). \\
%p_{cv}^{(0)}(\alpha) &= 1/q.
\end{split}
\]

Note also that, messages reaching symbol nodes of degree one or degree two check nodes do not participate in messages that are sent back from these nodes (see Figure~\ref{fig:messages}). Therefore, in a multiplicatively repeated non-binary LDPC code the decoder does not need to propagate messages either to those symbol nodes of degree one or to their adjacent check nodes of degree two (see the upper part of the Tanner graph in Figures~\ref{fig:example-c2} and~\ref{fig:example-c3}). Consequently, in the following steps 2 and 3 (that is, in the message-passing part of the algorithm), for decoding we only consider the lower part of the graphs, that is, the mother code $C_1$.
%after the variable nodes of degree 1 pass the initial messages to the upper part of the graph,

\emph{Step 2. Messages from checks to symbols:}
Each check node $z_m$ has incoming messages $q_{mn}^{(\ell)}$ received in the iteration $\ell$ from its adjacent symbol nodes $x_n$, for all $n \in \mathcal{N}(m)$. In the subsequent iteration, we calculate the messages sent back from the check node to its neighboring nodes, $r_{mn}^{(\ell+1)}$, but only using extrinsic information, that is, the message sent back from the check node $z_m$ to the symbol node $x_n$ is calculated using only the incoming messages from other edges, that is $q_{mn'}^{(\ell)}$ with $n' \in \mathcal{N}(m) \setminus \{n\}$.

Therefore, we compute\footnote{Note that, for convenience instead of computing $f(x) = h^{-1}(x)$ we could use $f(h(x)) = x$.} first:
\[
\tilde{q}_{mn}^{(\ell)}(\alpha) = q_{mn}^{(\ell)}(h_{mn}^{-1}\alpha), \quad \forall \alpha \in \operatorname{GF}(2^p)
\]
and then
\[
\tilde{r}_{mn}^{(\ell+1)}(\alpha) = \bigotimes_{\mathclap{n' \in \mathcal{N}(m) \setminus \{n\}}} \tilde{q}_{mn'}^{(\ell)}(\alpha), \quad \forall \alpha \in \operatorname{GF}(2^p)
\]
where $q_1 \otimes q_2$ is a convolution of $q_1$ and $q_2$, that is:
\[
(q_1 \otimes q_2)(\alpha) = \sum_{\mathclap{\substack{\alpha_1,\alpha_2 \in \operatorname{GF}(2^p) \\ \alpha=\alpha_1+\alpha_2}}} q_1(\alpha_1) q_2(\alpha_2), \quad \forall \alpha \in \operatorname{GF}(2^p).
\]

This is the most complex part of the decoding. However, the convolution can be efficiently calculated in the frequency domain using the Fourier transform over finite fields. This transform can be easily computed when the Galois field $\operatorname{GF}(q)$ is a binary extension field with order $q = 2^p$. The decoding of non-binary LDPC codes is then optimized using for instance the $p$-dimensional Walsh-Hadamard transform as proposed in \cite{Barnault_03, Declercq_07}. By applying the Walsh-Hadamard transform $\mathcal{W}\{\cdot\}$ the discrete convolution turns into a multiplication as follows:
\[
\mathcal{W}\{ \tilde{r}_{mn}^{(\ell+1)}(\alpha) \} = \prod_{\mathclap{n' \in \mathcal{N}(m) \setminus \{n\}}} \mathcal{W}\{ \tilde{q}_{mn'}^{(\ell)}(\alpha) \}, \quad \forall \alpha \in \operatorname{GF}(2^p).
\]

Moreover, since the Hadamard transform coincides with its inverse, it follows:
\[
\tilde{r}_{mn}^{(\ell+1)}(\alpha) = \mathcal{W}\{ \prod_{\mathclap{n' \in \mathcal{N}(m) \setminus \{n\}}} \mathcal{W}\{ \tilde{q}_{mn'}^{(\ell)}(\alpha) \} \}, \quad \forall \alpha \in \operatorname{GF}(2^p).
\]

Note that, both the input and output of the Walsh-Hadamard transform are $2^p$ dimensional real-valued vectors. Then, by $\mathcal{W}\{f(\alpha)\}$ we denote the result of applying the Walsh-Hadamard transform over the $2^p$ values of the function $f(\alpha)$, with $\alpha \in \operatorname{GF}(2^p)$, but considering only the component of the output vector that corresponds to the input $f(\alpha)$ for a given $\alpha$. This can be efficiently implemented via the fast Walsh-Hadamard transform.

Finally, each check node $z_m$ sends the message $r_{mn}^{(\ell+1)}$ to the symbol node $x_n$ for all $n \in \mathcal{N}(m)$ that is calculated as follows:
\[
r_{mn}^{(\ell+1)}(\alpha) = \tilde{r}_{mn}^{(\ell+1)}(h_{mn}\alpha), \quad \forall \alpha \in \operatorname{GF}(2^p).
\]

Note that, this is the channel coding version of the decoding algorithm, that is, when the transmitted message is always a codeword with zero syndrome. For the source coding with side information problem, the decoder look for the closest word with a given syndrome $z$ (the syndrome of $x$), and thus the decoding is calculated as follows:
\[
r_{mn}^{(\ell+1)}(\alpha) = \tilde{r}_{mn}^{(\ell+1)}(h_{mn}\alpha - z_m), \quad \forall \alpha \in \operatorname{GF}(2^p).
\]

\emph{Step 3. Messages from symbols to checks:}
Each symbol node $x_n$ has incoming messages $r_{mn}^{(\ell+1)}$ (computed in step~2) received from its adjacent check nodes $z_m$, for all $m \in \mathcal{M}(n)$. Then, messages $q_{mn}^{(\ell+1)}$ are sent back from the symbol node $x_n$ to its neighboring nodes, again only using extrinsic information, that is, the message sent back from the symbol node $x_n$ to the check node $z_m$ is calculated using only the incoming messages from other edges, that is $r_{m'n}^{(\ell+1)}$ with $m' \in \mathcal{M}(n) \setminus \{m\}$, as follows:
\[
q_{mn}^{(\ell+1)}(\alpha) = \beta_m\, p_{n}^{(0)}(\alpha) \prod_{\mathclap{m' \in \mathcal{M}(n) \setminus \{n\}}} r_{m'n}^{(\ell+1)}(\alpha), \quad \forall \alpha \in \operatorname{GF}(2^p)
\]
where $\beta_m$ is a normalizing factor such that messages $q_{mn}^{(\ell+1)}(\alpha)$ are probabilities, that is, $\beta_m$ is chosen such that:
\[
\sum_{\mathclap{\alpha \in \operatorname{GF}(2^p)}} q_{mn}^{(\ell+1)}(\alpha) = 1.
\]

\emph{Step 4. Tentative decision:}
Finally, for each symbol node in the mother code $C_1$ we calculate an estimation of the \emph{a posteriori} probability of symbol $x_n$ being $\alpha$, with $n = 1, \ldots, N$, as follows:
\[
p_n^{(\ell)}(\alpha) = p_n^{(0)} \prod_{\mathclap{m' \in \mathcal{M}(n)}} r_{m'n}^{(\ell)}(\alpha), \quad \forall \alpha \in \operatorname{GF}(2^p).
\]

We make then a tentative decision for the value of each symbol $x_n$ based on the highest probability:
\[
\hat{x}_n^{(\ell)} = \max_{\alpha \in \operatorname{GF}(2^p)} \{p_n^{(\ell)}(\alpha)\}.
\]

Multiplicative repetition symbols $x_{tN+n}$ of $x_n$, for each $n = 1, \ldots, N$, are then calculated with the newly obtained value $\hat{x}_n^{(\ell)}$ of symbol $x_n$ and using the parity-check constraints $x_{tN+n} = r_{tN+n} x_n$, for all $t=1,\ldots,T-1$.

Once we have an estimate of the values for all symbols, $\hat{x}$, we may calculate all the parity-check constraints to verify if they are satisfied (syndrome validation), that is, for the source coding with side information problem we verify if the received syndrome $z$ equals $H\hat{x}$. In such a case the algorithm concludes with successful decoding. Otherwise, the decoding algorithm continues iteratively, repeating steps 2 to 4, until the parity-check constraints are satisfied or a maximum number of iterations is reached without successful decoding. Note, however, that sometimes (typically in hardware implementations) the decoding algorithm continues iteratively without verifying whether the constraints are satisfied, that is, without a syndrome validation step. In that case the algorithm concludes after a given number of decoding iterations.

Compared to the decoding of binary LDPC codes, non-binary LDPC decoding demands a high computational complexity in the check-node processing (that is, when computing the messages from checks to symbols in Step~2 of the decoding algorithm) and requires a large amount of memory to store the messages exchanged in each iteration. However, there are several proposals in the literature to reduce both computational complexity and memory requirement.

Recent hardware (HW) implementations of non-binary LDPC codes for CV-QKD \cite{wei2025fpt} suggest that the constructions discussed here should similarly be amenable to HW. Let us review in more detail the state-of-the-art to understand where the challenges may lie. In \cite{Ferraz_22} there are summarized multiple HW implementations (in FPGA and ASIC architectures) of non-binary LDPC decoders. Some of these also consider the sum-product algorithm here described, over finite fields of order $2^8$, and with block-lengths of up to 1024 symbols and rate one half. We instead require codes of rate $1/3$, with less computational complexity. However, to the best of the authors’ knowledge, there are no hardware implementations of non-binary LDPC codes over finite fields of higher orders. In particular, its implementation over a finite field of order $2^{10}$ remains open.

\section{Results}
\label{sec:results}

Comprehensive simulations were performed to analyze the performance and efficiency of multiplicatively repeated non-binary LDPC codes, and such results are initially presented in Section~\ref{sec:efficiency}. Next, we verified that these codes are efficient enough to exchange secret-keys over long distances using a CV-QKD protocol, and secret-key rates are then given in Section~\ref{sec:secretkeyrate}.

\subsection{Performance and reconciliation efficiency}
\label{sec:efficiency}

In this section we study the performance and efficiency of multiplicatively repeated non-binary LDPC codes for correcting errors at very low SNRs, and compare these codes to other similar proposals in the literature. Simulations were performed over the binary input additive white Gaussian noise (BIAWGN) channel, as it is commonly used in prior work \cite{Jouguet_11, Jouguet_14b, Johnson_16, Shirvanimoghaddam_16, Bai_17, Milicevic_18, Zhou_19, Mani_21, Jeong_22, Yang_23}. This is a good model that approximates well, though not exactly, the correlations between input and output \cite{Laudenbach_18}. This channel is also the correct model for binary modulated CV-QKD \cite{Leverrier_23}. Although we could perform a similar analysis for this protocol, to avoid being repetitive, we focused on the application to Gaussian modulated CV-QKD.

\begin{figure}
\centering
\includegraphics[width=0.8\linewidth]{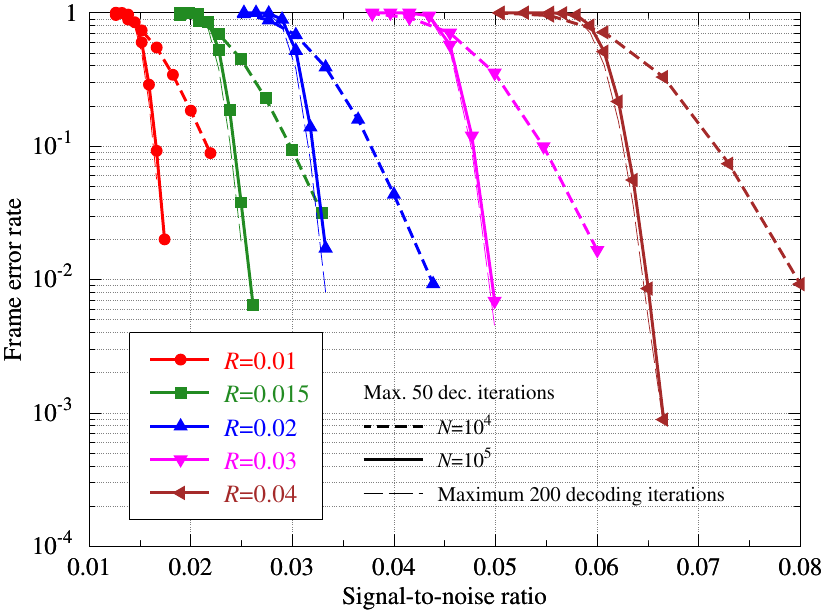}
\caption{Performance (that is, frame error rate as a function of signal-to-noise ratio) of low-rate multi-edge type LDPC codes over the BIAWGN channel, for several code rates $R$ and code lengths $N$ in bits.}
\label{fig:snr-met-ldpc}
\end{figure}

First, we have simulated the performance of low-rate multi-edge type LDPC codes proposed in~\cite{Richardson_04}. We designed ensembles of irregular multi-edge type LDPC codes using a modified version of the differential evolution algorithm described in~\cite{Shokrollahi_00}. The designed codes (see Table~\ref{tab:met-ldpc-ensembles}) have thresholds similar to others published in the literature. Thresholds were computed using a modified version of the discretized density evolution algorithm described in~\cite{Chung_01}. Both, differential and density evolution were modified according to the suggestions given in~\cite{Richardson_04, Rathi_05}, where the authors describe how the analysis for standard LDPC codes given in~\cite{Richardson_01a, Richardson_01b} extends to multi-edge type LDPC codes. Then, we constructed instances of the code ensembles given in Table~\ref{tab:met-ldpc-ensembles} using a modified progressive edge-growth algorithm \cite{Hu_05}. Code lengths of $10^4$ and $10^5$ bits were chosen\footnote{Note that, the code lengths here chosen are relatively short compared to the lengths considered in most proposals using multi-edge type LDPC codes in CV-QKD. For instance, a code length of $2^{20}$ bits was considered in~\cite{Jouguet_11}, a length of $10^6$ bits were used in~\cite{Wang_17, Wang_18, Milicevic_18}, and $1.024 \times 10^6$ bits length in~\cite{Mani_21}.}. Numerical results were finally computed using iterative LDPC decoding. For decoding we used a sum-product algorithm with serial schedule and a maximum of $50$ decoding iterations (that is, after each iteration we make a tentative decision and syndrome validation, thus the algorithm stops assuming that decoding was successful if the syndrome is satisfied or the maximum number of decoding iterations is reached). Figure~\ref{fig:snr-met-ldpc} shows the performance of these codes over the BIAWGN channel. For the codes of $10^5$ bits length, simulations were also performed increasing the maximum number of decoding iterations to $200$, but however, as shown, the performance does not improve significantly.

\begin{table}
\centering
\begin{tabular}{|c|c|ccc|c|ccc|}
\hline
$R$ & $\nu_m$ & \multicolumn{3}{|c|}{$m$} & $\mu_m$ & \multicolumn{3}{|c|}{$m$} \\
\hline
$0.01$ & $0.012$ & $2$ & $95$ & $0$ & $0.004$ & $4$ & $0$ & $0$ \\
       & $0.009$ & $3$ & $95$ & $0$ & $0.007$ & $5$ & $0$ & $0$ \\
       & $0.979$ & $0$ & $0$  & $1$ & $0.942$ & $0$ & $2$ & $1$ \\
       &         &     &      &     & $0.037$ & $0$ & $3$ & $1$ \\
\hline
\hline
$0.015$ & $0.028$ & $2$ & $53$ & $0$ & $0.005$ & $3$ & $0$ & $0$ \\
        & $0.009$ & $3$ & $53$ & $0$ & $0.017$ & $4$ & $0$ & $0$ \\
        & $0.963$ & $0$ & $0$  & $1$ & $0.928$ & $0$ & $2$ & $1$ \\
        &         &     &      &     & $0.035$ & $0$ & $3$ & $1$ \\
\hline
\hline
$0.02$ & $0.031$ & $2$ & $45$ & $0$ & $0.019$ & $4$ & $0$ & $0$ \\
       & $0.013$ & $3$ & $45$ & $0$ & $0.005$ & $5$ & $0$ & $0$ \\
       & $0.956$ & $0$ & $0$  & $1$ & $0.888$ & $0$ & $2$ & $1$ \\
       &         &     &      &     & $0.068$ & $0$ & $3$ & $1$ \\
\hline
\hline
$0.03$ & $0.017$ & $2$ & $58$ & $0$ & $0.011$ & $9$ & $0$ & $0$ \\
       & $0.025$ & $3$ & $58$ & $0$ & $0.001$ & $10$ & $0$ & $0$ \\
       & $0.958$ & $0$ & $0$  & $1$ & $0.438$ & $0$ & $2$ & $1$ \\
       &         &     &      &     & $0.52$  & $0$ & $3$ & $1$ \\
\hline
\hline
$0.04$ & $0.027$ & $2$ & $44$ & $0$ & $0.015$ & $9$ & $0$ & $0$ \\
       & $0.029$ & $3$ & $44$ & $0$ & $0.001$ & $6$ & $0$ & $0$ \\
       & $0.944$ & $0$ & $0$  & $1$ & $0.368$ & $0$ & $2$ & $1$ \\
       &         &     &      &     & $0.576$ & $0$ & $3$ & $1$ \\
\hline
\end{tabular}
\caption{Multi-edge type LDPC ensembles.}
\label{tab:met-ldpc-ensembles}
\end{table}

\begin{figure}
\centering
\includegraphics[width=0.8\linewidth]{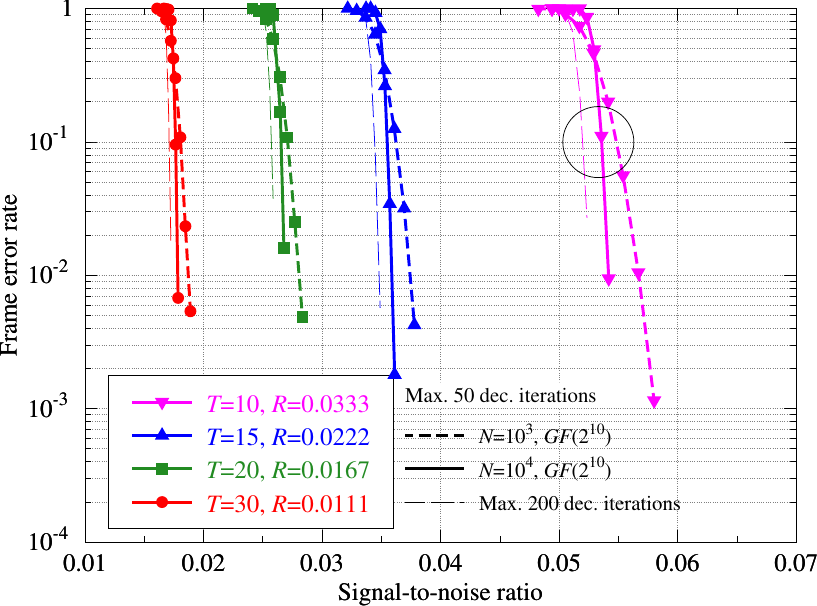}
\caption{Performance of multiplicatively repeated non-binary LDPC codes over $\operatorname{GF}(2^{10})$ and the BIAWGN channel, for several code rates $R$ and code lengths $N$ in symbols.}
\label{fig:snr-multirep}
\end{figure}

Next, we have simulated the performance of multiplicatively repeated non-binary LDPC codes. We considered as mother code $C_1$ a $(2,3)$-regular non-binary LDPC code over $\operatorname{GF}(2^{10})$ of rate $1/3$. From such a code we constructed multiplicatively repeated codes of lower rates $1/30$, $1/45$, $1/60$, and $1/90$, as described in Section~\ref{sec:multirep}. For the mother code two code lengths of $N=10^3$ and $10^4$ symbols, were considered. Given that each element of the finite field is represented by a binary polynomial of degree less than or equal to $9$ (that is, a polynomial with $10$ binary coefficients or $10$-bit string), we are using $10$ bits per symbol. Thus, the lengths considered for the mother codes equal the lengths in bits previously considered for multi-edge type LDPC codes. Note that, the length of a multiplicatively repeated code actually depends on the repetition parameter $T$, that is, the actual number of symbols or codeword length is $NT$. However, as we already argued in Section~\ref{sec:decoding}, multiplicative repetition symbols do not contribute to the messages computed in the message-passing part of the decoding algorithm. Therefore, for the sake of convenience, in the following when we refer to the length of multiplicatively repeated non-binary LDPC codes we consider the length $N$ of the mother code (instead of the actual but unrealistic code length $NT$). Numerical results were computed using iterative LDPC decoding. For decoding we used the sum-product algorithm described in Section~\ref{sec:decoding} with a maximum of $50$ decoding iterations (again, making a tentative decision and verifying the syndrome after each iteration). Figure~\ref{fig:snr-multirep} shows the performance of these multiplicatively repeated non-binary LDPC codes over the BIAWGN channel. As shown, it is noteworthy that the performance of shorter codes, of $10^3$ symbol lengths, is almost as good as that of longer codes. For the codes of $10^4$ symbols length, simulations were also performed increasing the maximum number of decoding iterations to $200$. As shown, unlike multi-edge type LDPC codes, the number of iterations plays a determining role for multiplicatively repeated non-binary LDPC codes. This behavior is depicted in Figure~\ref{fig:snr-multirep}, where a circle marks the performance (at a frame error rate of $10^{-1}$) of the code of rate $R=1/30$ considering different code lengths and decoding iterations. In summary, to improve the performance (and consequently the efficiency) of these codes, it is necessary to increase both the code length (as usual) and the maximum number of decoding iterations.

\begin{table}
\centering
\begin{tabular}{|c|c|c|c|c|c|c|c|}
\hline
\multicolumn{4}{|c|}{Multi-edge type} & \multicolumn{4}{|c|}{Multiplicatively repeated} \\
\hline
$R$ & $N$ & iters & $\beta$ & $R$ & $N$ & iters & $\beta$ \\
\hline
$0.01$  & $10^5$ & $50$  & $0.8475$ & $0.0111$ & $10^3$ & $200$  & $\bf 0.8732$ \\
$0.01$  & $10^5$ & $200$ & $0.875$  & $0.0111$ & $10^4$ & $200$ & $\bf 0.9079$ \\
$0.015$ & $10^5$ & $50$  & $0.8647$ & $0.0166$ & $10^3$ & $200$  & $\bf 0.876$ \\
$0.015$ & $10^5$ & $200$ & $0.871$  & $0.0166$ & $10^4$ & $200$ & $\bf 0.9087$ \\
$0.02$  & $10^5$ & $50$  & $\bf 0.8793$ & $0.0222$ & $10^3$ & $200$  & $0.8775$ \\
$0.02$  & $10^5$ & $200$ & $0.8911$ & $0.0222$ & $10^4$ & $200$ & $\bf 0.9092$ \\
$0.03$  & $10^5$ & $50$  & $\bf 0.894$  & $0.0333$ & $10^3$ & $200$  & $0.8781$ \\
$0.03$  & $10^5$ & $200$ & $0.898$  & $0.0333$ & $10^4$ & $200$ & $\bf 0.9112$ \\
\hline
\end{tabular}
\caption{Information reconciliation efficiencies.}
\label{tab:efficiency}
\end{table}

In the following we study and compare the efficiency of multiplicatively repeated non-binary LDPC codes with other proposals, but let us first see how we calculate it. Let $R$ be the rate of the code used for correcting errors, then the reconciliation efficiency in CV-QKD, denoted by $\beta$, is calculated as follows:
\begin{equation}
\label{eq:efficiency}
\beta = \frac{R}{C},
\end{equation}
where $C$ is the channel capacity, that is, here the capacity of the BIAWGN channel. Hence, the channel capacity and therefore also the efficiency are functions of SNR. Note that, for small SNR values $s$ the capacity of the BIAWGN channel is well approximated by that of the AWGN channel, given by $C = \frac{1}{2} \log_2(1+s)$.

Table~\ref{tab:efficiency} shows the reconciliation efficiencies, $\beta$, for the multi-edge type LDPC and multiplicatively repeated non-binary LDPC codes simulated in Figures~\ref{fig:snr-met-ldpc} and~\ref{fig:snr-multirep}, respectively. A number of significant cases were chosen for several code rates $R$, code lengths $N$, and maximum number of decoding iterations (iters). In the case of multi-edge type LDPC codes were only considered larger code lengths of $10^5$ bits, but results are compared for a maximum of $50$ and $200$ decoding iterations. For the multiplicatively repeated non-binary LDPC codes were considered both code lengths, $10^3$ and $10^4$ symbols, but only a maximum of $200$ decoding iterations. Efficiencies were calculated for a target frame error rate (FER) of $10^{-1}$ as suggested in \cite{Martinez_13}, that is, given a code of rate $R$ we first compute the SNR for which the code works at a FER of $10^{-1}$, then we calculate the channel capacity $C$ and efficiency $\beta$ using equation~\eqref{eq:efficiency} with the obtained SNR. As shown in Table~\ref{tab:efficiency}, multiplicatively repeated non-binary LDPC (even when using short block-length) codes outperform multi-edge type LDPC codes, but the maximum number of decoding iterations is a determining parameter, since, while multi-edge type LDPC codes do not substantially improve the efficiency with larger decoding iterations, this is not the case with multiplicatively repeated non-binary LDPC codes. A comprehensive analysis of this behavior was performed below.

\begin{figure}
\centering
\includegraphics[width=0.8\linewidth]{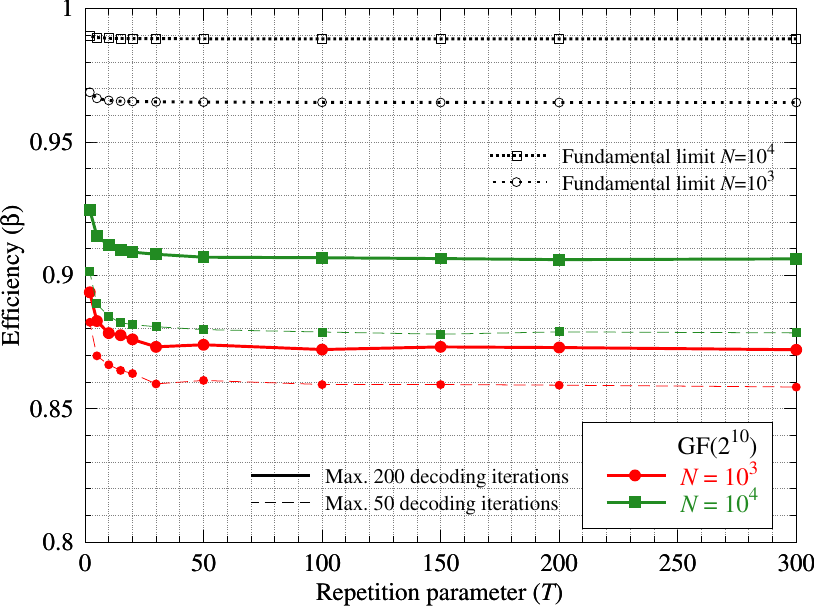}
\caption{Reconciliation efficiency of multiplicatively repeated non-binary LDPC codes over $\operatorname{GF}(2^{10})$. Efficiency was computed considering a target FER of $10^{-1}$.}
\label{fig:efficiency_t}
\end{figure}

Figure~\ref{fig:efficiency_t} shows the reconciliation efficiency of multiplicatively repeated non-binary LDPC codes over $\operatorname{GF}(2^{10})$ as a function of the repetition parameter $T$. As shown, the efficiency gradually decreases at the beginning, that is with $2 \leq T \leq 10$, but surprisingly remains almost constant for higher values of the repetition parameter, that is for $T \geq 20$. Therefore, low-rate and very low-rate codes are almost equally efficient. It is noteworthy that extremely low-rate codes, that is, codes of rates $R=1/300$, $R=1/600$ and up to $R=1/900$, with length $10^4$ symbols, have an efficiency above $90\%$. Furthermore, for very low-rate codes, that is $R<0.02$, the efficiency of codes of length $10^3$ symbols is even better than the efficiency of multi-edge type LDPC codes of length $10^5$, as reported in Table~\ref{tab:efficiency}. Note also that, to better understand the significance of the number of decoding iterations, the figure shows simulation results for a maximum of $50$ (dashed line) and $200$ (solid line) iterations.

In the figure, we also show the fundamental limits on the efficiency when reconciling errors using multiplicatively repeated non-binary LDPC codes over $\operatorname{GF}(2^{10})$ with a mother code of lengths $N=10^3$ and $N=10^4$. We used recent results in non-asymptotic classical information theory \cite{Tomamichel_17} for upper bounding the reconciliation efficiency in CV-QKD with only one-way communications. These results were adapted to calculate the efficiency $\beta$ as given in eq.~\eqref{eq:efficiency}:
\begin{equation}
\beta(n, \epsilon, \sigma) = 1 - \frac{\sqrt{v(\sigma) / n}}{1 - h(\sigma)} \Phi^{-1}(1 - \epsilon),
\end{equation}
where $n$ is the length of the code (in bits), $\epsilon$ is the frame error rate, and $\sigma$ is the signal-to-noise ratio. On the other hand, $\Phi(\cdot)$ is the cumulative standard normal distribution, $h(\sigma) = 1 - C$ is the conditional entropy, and $v(\sigma) = e(\sigma) - h(\sigma)^2$ is the conditional entropy variance, where:
\[
f_{XY}(x,y) = \sqrt{\frac{\sigma}{8\pi}} e^{\sigma (y-x)^2 / 2}, \qquad
f_Y(y) = f_{XY}(1,y) + f_{XY}(-1,y),
\]
\[
C = - \int_{-\infty}^{\infty} f_Y(y) \log_2 f_Y(y) dy + \frac{1}{2} \log_2 \left( \frac{\sigma}{2\pi e} \right) \approx \frac{1}{2} \log_2 (1 + \sigma),
\]
and
\[
e(\sigma) = 2 \int_{-\infty}^{\infty} f_{XY}(1,y) \left( \log_2 \frac{f_{XY}(1,y)}{f_Y(y)} \right)^2 dy.
\]

Note that each point in the figure corresponds to the efficiency of a multiplicatively repeated non-binary LDPC code for a given repetition parameter $T$, where each symbol of the mother code is then multiplicatively repeated $T$ times, thus expanding the codeword length to $NT$ symbols. Given that we are using 10 bits per symbol, the length of the equivalent binary code is $n = 10 NT$ bits. Furthermore, for each $T$ value we consider the signal-to-noise ratio $\sigma$ at which the efficiency shown was achieved. Finally, taking into account that efficiencies were calculated for a target frame error rate of $\epsilon = 10^{-1}$, we may obtain an upper bound for the reconciliation efficiency given by $\beta(n, \epsilon, \sigma)$.

For a constant block-length code, this upper bound on the reconciliation efficiency is a monotonically decreasing function for decreasing signal-to-noise ratios. However, both the fundamental limits and reconciliation efficiencies shown remain constant. This behavior occurs because, as the repetition parameter increases, more symbol nodes are multiplicatively repeated, thereby increasing the effective length $n$ (in bits) of the reconciled string.

\begin{figure}
\centering
\includegraphics[width=0.8\linewidth]{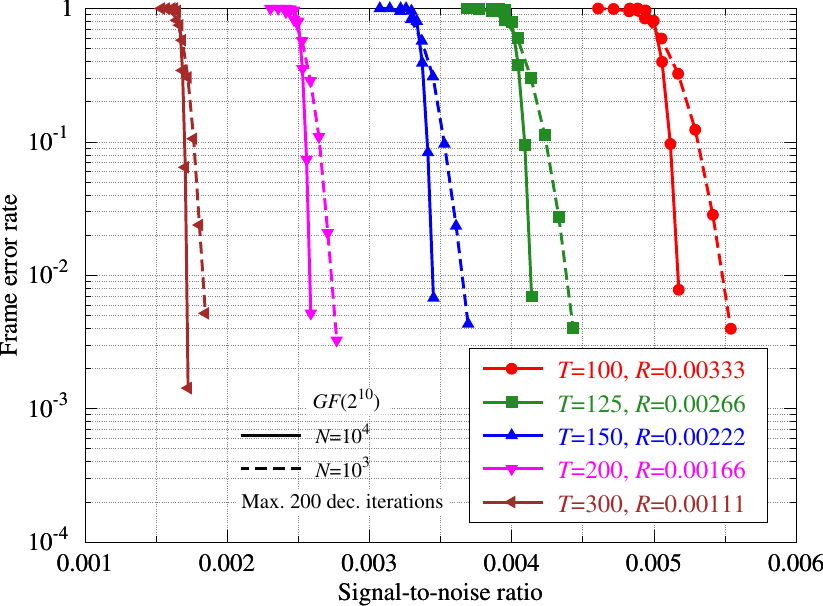}
\caption{Performance of multiplicatively repeated non-binary LDPC codes over $\operatorname{GF}(2^{10})$ and the BIAWGN channel, for several ultra-low code rates $R$ and code lengths $N$ in symbols.}
\label{fig:snr-multirep2}
\end{figure}

Figure~\ref{fig:snr-multirep2} shows again the performance of multiplicatively repeated non-binary LDPC codes but now for ultra-low rates, up to $R=0.00111$ (that is, with repetition parameter $T=300$). The mother codes $C_1$ used are the same as for Figure~\ref{fig:snr-multirep}, that is, a $(2,3)$-regular non-binary LDPC code over $\operatorname{GF}(2^{10})$ of rate $1/3$ and lengths of $10^3$ and $10^4$ symbols. However, in this case simulations were performed only considering $200$ decoding iterations maximum. The figure shows the performance of those codes with lowest code rates that perform well over the BIAWGN channel (as confirmed in Figure~\ref{fig:efficiency_t}).

\begin{figure}
\centering
\includegraphics[width=0.8\linewidth]{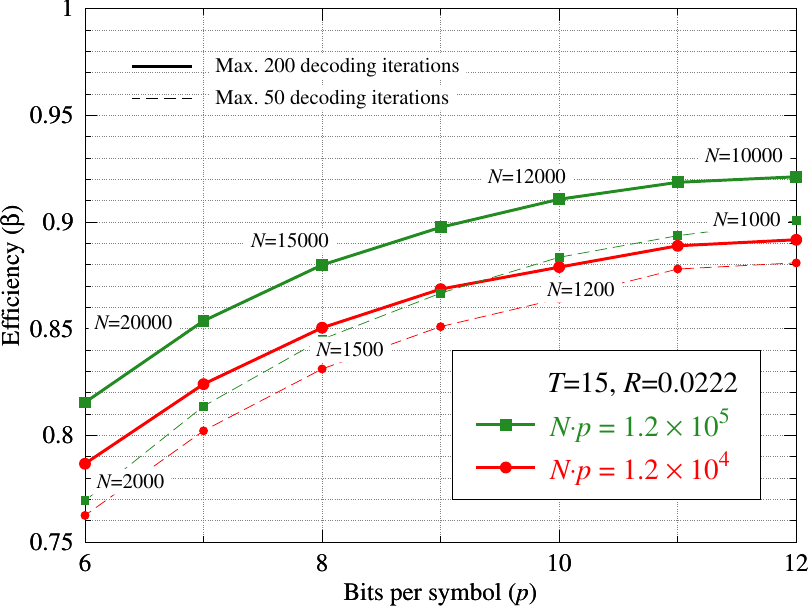}
\caption{Reconciliation efficiency of multiplicatively repeated non-binary LDPC codes over $\operatorname{GF}(2^p)$ with repetition parameter $T=15$. Efficiency was computed considering a target FER of $10^{-1}$.}
\label{fig:efficiency_q}
\end{figure}

Finally, we analyze the relationship between the efficiency and the order of the Galois field used in the construction and decoding of multiplicatively repeated non-binary LDPC codes. Figure~\ref{fig:efficiency_q} reports the reconciliation efficiency of multiplicatively repeated non-binary LDPC codes over different Galois fields $\operatorname{GF}(q)$ being a binary extension field with order $q = 2^p$. For all the codes the repetition parameter is $T=15$, hence the code rate is $R=1/45$. However, different code length $N$ were chosen such that the mother code lengths in bits are $Np = 1.2 \times 10^4$ and $Np = 1.2 \times 10^5$. As shown, as the order of the Galois fields increases, the efficiency also improves. Therefore, high orders are also a necessary condition to achieve good performance and reconciliation efficiency using the proposed codes. Again, as in the previous figure, results are shown for a maximum of $50$ (dashed line) and $200$ (solid line) decoding iterations.

\subsection{Secret-key rate}
\label{sec:secretkeyrate}

The asymptotic secret-key rate for collective attacks of a CV-QKD protocol using reverse reconciliation is given by $K = \beta I_{\rm AB} - \chi_{\rm BE}$, where $I_{\rm AB}$ is the mutual information between Alice and Bob (emitter/decoder and receiver/encoder, respectively), $\chi_{\rm BE}$ is the Holevo bound on the information leaked to the eavesdropper Eve (that is, the maximum information she may have access to) for reverse reconciliation \cite{Lodewyck_07}, and $\beta$ is the reconciliation efficiency. This efficiency gives a fraction of the raw keys shared by Alice and Bob after the reconciliation procedure, that is, the length $n \beta I_{\rm AB}$ of the reconciled bit strings that the parties are left with. Good efficiency values are then necessary to achieve high secret-key rates over long distances, but there are also other parameters that need to be considered, such as FER, since both efficiency and FER are correlated as shown in Section~\ref{sec:efficiency}.

In the finite-size scenario, that is, when considering finite-size effects (mainly in the parameter estimation procedure) the secret-key rate is then given by \cite{Leverrier_10}
\begin{equation}
\label{eq:secret-key-rate}
K = \frac{n}{N} (1-F) (\beta I_{\rm AB} - \chi_{\rm BE} - \Delta(n)),
\end{equation}
where $n$ is the length of the raw key (a reconciled and therefore common bit string) used in the privacy amplification procedure, $N$ is the number of exchanged signals (thus, $N-n$ signals are used for parameter estimation), $F$ is the reconciliation FER, and $\Delta(n)$ is a function related to the security of privacy amplification in the finite-size regime. When $n \geq 10^4$ this function is essentially determined by \cite{Leverrier_10}
\[
\Delta(n) \approxeq 7 \sqrt{\frac{\log_2(2/\bar{\epsilon})}{n}},
\]
where $\bar{\epsilon}$ is a smoothing parameter.

\begin{figure}
\centering
\includegraphics[width=0.8\linewidth]{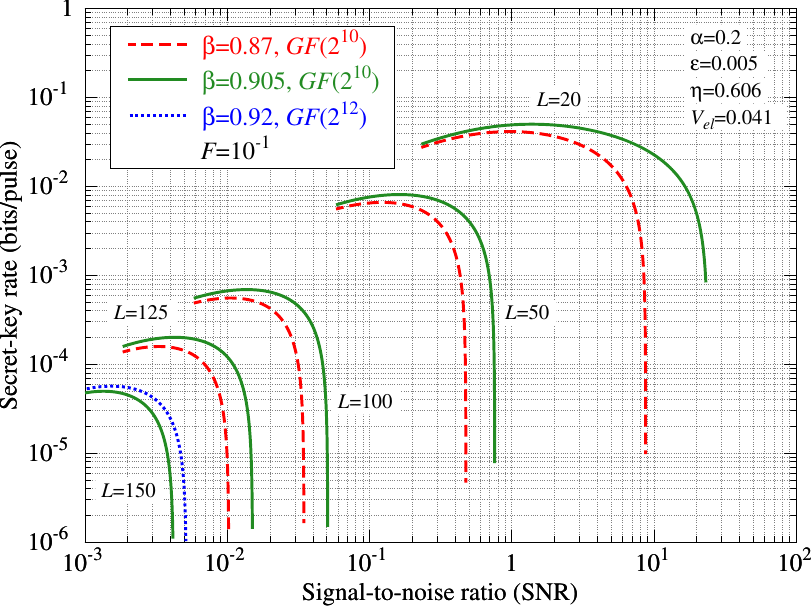}
\caption{Secret-key rate for collective attacks with respect to the SNR.}
\label{fig:skr-snr}
\end{figure}

Figure~\ref{fig:skr-snr} shows the finite-size secret-key rate as a function of the signal-to-noise ratio for several transmission distances $L=20$, $L=50$, $L=100$, $L=125$ and $L=150$, and different reconciliation efficiencies $\beta = 0.87$, $\beta = 0.9$ and $\beta = 0.92$. Efficiencies $\beta = 0.87$ and $\beta = 0.9$ correspond to multiplicatively repeated non-binary LDPC codes of length $10^3$ and $10^4$, respectively, over $\operatorname{GF}(2^{10})$, as reported in Table~\ref{tab:efficiency}. An efficiency of $\beta = 0.92$ corresponds to multiplicatively repeated non-binary LDPC codes of length $10^4$ over $\operatorname{GF}(2^{12})$. For the secret-key rate calculation, the quantum channel and CV-QKD devices were characterized using common parameters previously published in the literature \cite{Lodewyck_07, Milicevic_18, Jeong_22, Yang_23}. Hence, we assume the standard loss of a single-mode optical fiber of $\alpha = 0.2$~dB/km, a constant excess channel noise of $\varepsilon = 0.005$ (in shot noise units), and Bob’s homodyne detector efficiency of $\eta = 0.606$, with electronic noise $V_{\rm el} = 0.041$ (in shot noise units). Alice’s modulation variance $V_A$ (in shot noise units) is considered within the interval $[1,100]$ and optimized at each transmission distance to maximize the secret-key rate. Furthermore, as suggested in~\cite{Leverrier_10, Milicevic_18} we have also considered a raw key length of $n = 10^{12}$ bits with $N = 2n$, and a conservative choice for the security parameter $\bar{\epsilon} = 10^{-10}$.

\begin{figure}
\centering
\includegraphics[width=0.8\linewidth]{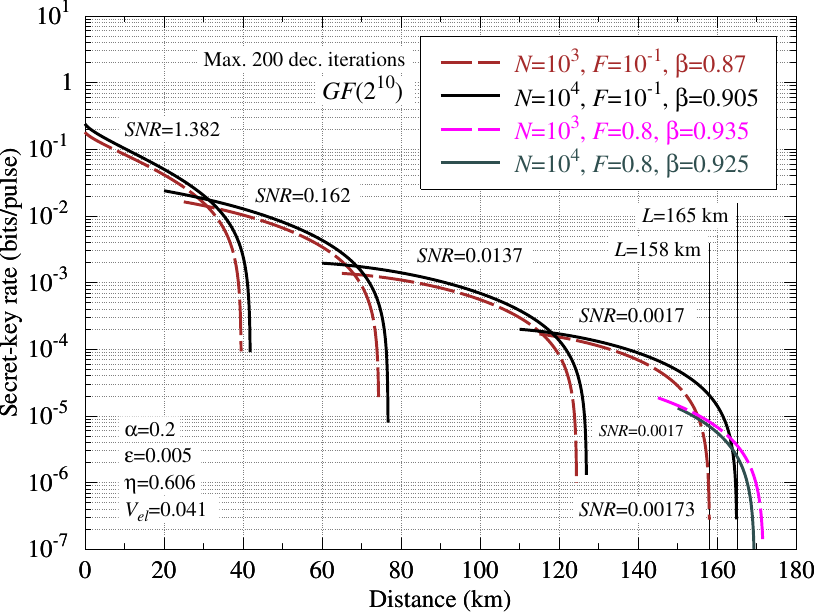}
\caption{Secret-key rate for collective attacks with respect to the distance.}
\label{fig:skr-distance}
\end{figure}

Finally, Figure~\ref{fig:skr-distance} shows the finite-size secret-key rate with respect to the transmission distance using the optimal SNR values for each distance, as calculated for Figure~\ref{fig:skr-snr}. We have considered the performance and efficiency of multiplicatively repeated non-binary LDPC codes over $\operatorname{GF}(2^{10})$ of lengths $N=10^3$ symbols (long-dashed lines) and $N=10^4$ symbols (solid lines), that is, as previously reported in Table~\ref{tab:efficiency}, we considered the reconciliation efficiencies of $\beta=0.87$ and $\beta=0.90$, respectively, with a FER of $10^{-1}$. As shown, the maximum achievable distance of a CV-QKD protocol using the proposed codes is around $158$~km with a multiplicatively repeated non-binary LDPC code over $\operatorname{GF}(2^{10})$ of $10^3$ symbols length, and around $165$~km with a code of $10^4$ symbols length, in both cases with a FER of $F=10^{-1}$. Additionally, we have also considered an efficiency of $\beta=0.925$ and $\beta=0.935$, which are achieved at a higher FER of $80\%$, both for the codes of length $10^4$ symbols and $10^3$ symbols, respectively. The maximum distance is then slightly increased to approximately $169.4$~km using the code of length $10^4$ symbols, and $171.7$~km using the code of length $10^3$ symbols.

\begin{figure}
\centering
\includegraphics[width=0.8\linewidth]{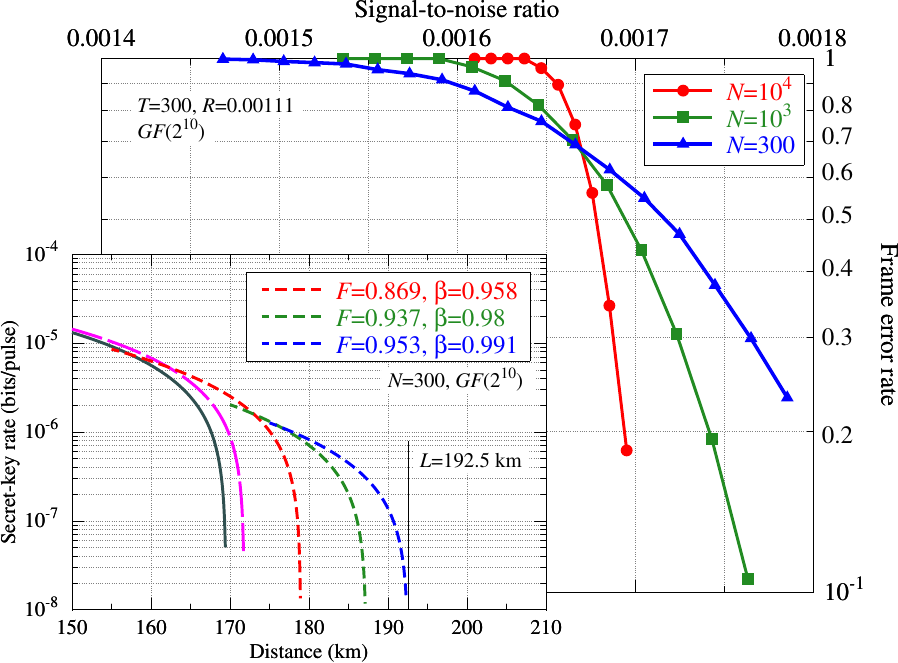}
\caption{Performance of multiplicatively repeated non-binary LDPC codes for several code lengths $N$, and secret-key rate for collective attacks with respect to the distance using short block-length codes.}
\label{fig:skr-length}
\end{figure}

Regarding the performance of error-correcting codes in the high FER region, in coding theory it is well known that in this region there is an SNR value at which codes of different block-lengths have the same performance. Then, on the one hand, for higher SNRs the performance of long codes is much better than that of short ones. On the other hand, however, for lower SNRs the performance of short block-length codes is better. This behavior is shown in Figure~\ref{fig:skr-length} (although this can also be seen in Figures~\ref{fig:snr-met-ldpc}, \ref{fig:snr-multirep} and~\ref{fig:snr-multirep2}). According to this behavior, it is interesting to study how short block-length codes can help to increase the maximum distance at which a key can be securely exchanged and reconciled. Indeed, Figure~\ref{fig:skr-length} also shows that for short codes there is still a range of SNRs for which it is still possible to exchange secret-keys, and thus increase the maximum secure distance of a CV-QKD protocol.

\begin{figure}
\centering
\includegraphics[width=0.8\linewidth]{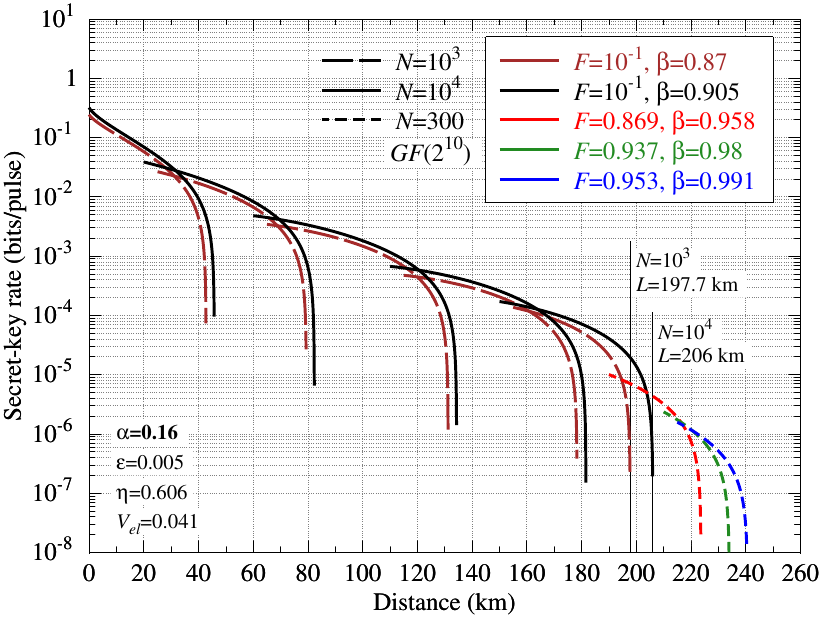}
\caption{Secret-key rate for collective attacks with respect to the distance using an ultra-low loss fiber.}
\label{fig:skr-distance2}
\end{figure}

In addition, to make this proposal comparable to other results reported in the literature, we included an additional figure. Figure~\ref{fig:skr-distance2} shows the secret-key rate again for collective attacks and the finite-size case, but considering the transmission over an ultra-low loss fiber with attenuation of $\alpha=0.16$~dB/km, such as in~\cite{ZhangYichen_20}. The secret-key rate was calculated considering the same codes, their performance and efficiency, as in Figures~\ref{fig:skr-distance} and~\ref{fig:skr-length}.

\section{Conclusions}
\label{sec:conclusions}

Multiplicatively repeated non-binary LDPC codes over a finite field of characteristic two were considered for correcting errors in the low SNR regime. These codes are of particular interest for information reconciliation in CV-QKD, since they outperform multi-edge type LDPC codes that were thought to be the best method for low-rate coding. The construction of these codes is very simple, and there is no need to design codes of different rates. Only a regular non-binary LDPC code, used as mother code, is required. Lower rate codes are then constructed from this mother code. They are also inherently rate-adaptive, which allows for an improved reconciliation efficiency when the channel parameter is estimated, as is the case with QKD. Furthermore, it has been shown that these codes perform well even for short code lengths, and decoding has also been shown to perform with almost the same computational complexity as that of the mother code, thus making them suitable for hardware implementations. The only comparative drawback is the larger number of decoding iterations needed to ensure good efficiency.

As shown, the proposed codes are able to distill secret-keys from a single mother code of short and intermediate block-length, that by multiplicatively repeating symbols spans nearly the whole SNR range, that is, most distances. Additionally, very short block-length codes working in the high FER regime can likewise be used to distill secret-keys particularly for longer distances.

\section*{Declarations}

\subsection*{Availability of data and materials}
Data sets generated during the current study are available from the corresponding author on reasonable request.

\subsection*{Competing interests}
The authors have no competing interests, or other interests that might be perceived to influence the results and/or discussion reported in this paper.

\subsection*{Funding}
This research has been partially supported by the Ministerio de Ciencia e Innovaci\'{o}n (MICINN), Government of Spain (grant PID2021-122905NB-C22).
This work was partially supported by Japan's Council for Science, Technology and Innovation (CSTI) under the Cross-ministerial Strategic Innovation Promotion Program (SIP) for ``Promoting the application of advanced quantum technology platforms to social issues'' (grant JPJ012367).

\subsection*{Authors' contributions}
J.M. constructed the multiplicatively repeated non-binary LDPC codes and decoder, and performed the simulations. J.M. wrote the first version of the manuscript. J.M. and D.E. reviewed the main manuscript text and discussed the simulation results. J.M. and D.E. supervised this work.

\bibliographystyle{unsrtnat}
\bibliography{low-rate_qkd}

\begin{thebibliography}{67}
\providecommand{\natexlab}[1]{#1}
\providecommand{\url}[1]{\texttt{#1}}
\expandafter\ifx\csname urlstyle\endcsname\relax
  \providecommand{\doi}[1]{doi: #1}\else
  \providecommand{\doi}{doi: \begingroup \urlstyle{rm}\Url}\fi

\bibitem[Gisin et~al.(2002)Gisin, Ribordy, Tittel, and Zbinden]{Gisin_02}
Nicolas Gisin, Gr\'{e}goire Ribordy, Wolfgang Tittel, and Hugo Zbinden.
\newblock Quantum cryptography.
\newblock \emph{Rev. Mod. Phys.}, 74\penalty0 (1):\penalty0 145--195, 2002.
\newblock \doi{10.1103/RevModPhys.74.145}.

\bibitem[Brassard and Salvail(1994)]{Brassard_94}
Gilles Brassard and Louis Salvail.
\newblock Secret-key reconciliation by public discussion.
\newblock In \emph{Eurocrypt'93, Workshop on the theory and application of
  cryptographic techniques on Advances in cryptology}, volume 765 of
  \emph{Lecture Notes in Computer Science}, pages 410--423, New York, 1994.
  Springer-Verlag.

\bibitem[Bennett et~al.(1988)Bennett, Brassard, and Roberts]{Bennett_88}
Charles~H. Bennett, Gilles Brassard, and Jean-Marc Roberts.
\newblock Privacy amplification by public discussion.
\newblock \emph{SIAM J. Comput.}, 17\penalty0 (2):\penalty0 210--229, 1988.
\newblock \doi{10.1137/0217014}.

\bibitem[Bennett and Brassard(2014)]{Bennett_84}
Charles~H. Bennett and Gilles Brassard.
\newblock Quantum cryptography: Public key distribution and coin tossing.
\newblock \emph{Theor. Comput. Sci.}, 560:\penalty0 7--11, 2014.
\newblock \doi{10.1016/j.tcs.2014.05.025}.

\bibitem[Elkouss et~al.(2011)Elkouss, Martinez-Mateo, and Martin]{Elkouss_11}
David Elkouss, Jesus Martinez-Mateo, and Vicente Martin.
\newblock Information reconciliation for quantum key distribution.
\newblock \emph{Quantum Inform. Comput.}, 11\penalty0 (3\&4):\penalty0
  226--238, 2011.
\newblock \doi{10.26421/QIC11.3-4-3}.

\bibitem[Martinez-Mateo et~al.(2012)Martinez-Mateo, Elkouss, and
  Martin]{Martinez_12}
Jesus Martinez-Mateo, David Elkouss, and Vicente Martin.
\newblock Blind reconciliation.
\newblock \emph{Quantum Inform. Comput.}, 12\penalty0 (9\&10):\penalty0
  791--812, 2012.
\newblock \doi{10.26421/QIC12.9-10-5}.

\bibitem[Martinez-Mateo et~al.(2013)Martinez-Mateo, Elkouss, and
  Martin]{Martinez_13}
Jesus Martinez-Mateo, David Elkouss, and Vicente Martin.
\newblock Key reconciliation for high performance quantum key distribution.
\newblock \emph{Sci. Rep.}, 3\penalty0 (1576), 2013.
\newblock \doi{10.1038/srep01576}.

\bibitem[Tarable et~al.(2024)Tarable, Paganelli, and Ferrari]{Tarable_24}
Alberto Tarable, Rudi~Paolo Paganelli, and Marco Ferrari.
\newblock Rateless protograph ldpc codes for quantum key distribution.
\newblock \emph{IEEE Transactions on Quantum Engineering}, 5:\penalty0 1--11,
  2024.
\newblock \doi{10.1109/TQE.2024.3361810}.

\bibitem[Jouguet and Kunz-Jacques(2014)]{Jouguet_14}
Paul Jouguet and Sebastien Kunz-Jacques.
\newblock High performance error correction for quantum key distribution using
  polar codes.
\newblock \emph{Quantum Inform. Comput.}, 14\penalty0 (3\&4):\penalty0
  329--338, 2014.
\newblock \doi{10.26421/QIC14.3-4-8}.

\bibitem[Martinez-Mateo et~al.(2015)Martinez-Mateo, Pacher, Peev, Ciurana, and
  Martin]{Martinez_15}
J.~Martinez-Mateo, C.~Pacher, M.~Peev, A.~Ciurana, and V.~Martin.
\newblock Demystifying the information reconciliation protocol {Cascade}.
\newblock \emph{Quantum Inform. Comput.}, 15\penalty0 (5\&6):\penalty0
  453--477, 2015.
\newblock \doi{10.26421/QIC15.5-6-6}.

\bibitem[Pacher et~al.(2015)Pacher, Grabenweger, Martinez-Mateo, and
  Martin]{Pacher_15}
Christoph Pacher, Philipp Grabenweger, Jesus Martinez-Mateo, and Vicente
  Martin.
\newblock An information reconciliation protocol for secret-key agreement with
  small leakage.
\newblock In \emph{2015 IEEE International Symposium on Information Theory
  (ISIT)}, pages 730--734, 2015.
\newblock \doi{10.1109/ISIT.2015.7282551}.

\bibitem[Van~Assche et~al.(2004)Van~Assche, Cardinal, and Cerf]{VanAssche_04}
G.~Van~Assche, J.~Cardinal, and N.J. Cerf.
\newblock Reconciliation of a quantum-distributed gaussian key.
\newblock \emph{IEEE Trans. Inf. Theory}, 50\penalty0 (2):\penalty0 394--400,
  2004.
\newblock \doi{10.1109/TIT.2003.822618}.

\bibitem[Leverrier et~al.(2008)Leverrier, All\'eaume, Boutros, Z\'emor, and
  Grangier]{Leverrier_08}
Anthony Leverrier, Romain All\'eaume, Joseph Boutros, Gilles Z\'emor, and
  Philippe Grangier.
\newblock Multidimensional reconciliation for a continuous-variable quantum key
  distribution.
\newblock \emph{Phys. Rev. A}, 77:\penalty0 042325, 2008.
\newblock \doi{10.1103/PhysRevA.77.042325}.

\bibitem[Jouguet et~al.(2011)Jouguet, Kunz-Jacques, and Leverrier]{Jouguet_11}
Paul Jouguet, S\'ebastien Kunz-Jacques, and Anthony Leverrier.
\newblock Long-distance continuous-variable quantum key distribution with a
  gaussian modulation.
\newblock \emph{Phys. Rev. A}, 84:\penalty0 062317, 2011.
\newblock \doi{10.1103/PhysRevA.84.062317}.

\bibitem[Jouguet et~al.(2014)Jouguet, Elkouss, and Kunz-Jacques]{Jouguet_14b}
Paul Jouguet, David Elkouss, and S\'ebastien Kunz-Jacques.
\newblock High-bit-rate continuous-variable quantum key distribution.
\newblock \emph{Phys. Rev. A}, 90:\penalty0 042329, 2014.
\newblock \doi{10.1103/PhysRevA.90.042329}.

\bibitem[Bai et~al.(2017)Bai, Yang, and Li]{Bai_17}
Zengliang Bai, Shenshen Yang, and Yongmin Li.
\newblock High-efficiency reconciliation for continuous variable quantum key
  distribution.
\newblock \emph{Jpn. J. Appl. Phys.}, 56\penalty0 (4):\penalty0 044401, 2017.
\newblock \doi{10.7567/JJAP.56.044401}.

\bibitem[Wang et~al.(2018)Wang, Zhang, Yu, and Guo]{Wang_18}
Xiangyu Wang, Yichen Zhang, Song Yu, and Hong Guo.
\newblock High speed error correction for continuous-variable quantum key
  distribution with multi-edge type {LDPC} code.
\newblock \emph{Sci. Rep.}, 8\penalty0 (10543), 2018.
\newblock \doi{10.1038/s41598-018-28703-4}.

\bibitem[Zhang et~al.(2020)Zhang, Chen, Pirandola, Wang, Zhou, Chu, Zhao, Xu,
  Yu, and Guo]{ZhangYichen_20}
Yichen Zhang, Ziyang Chen, Stefano Pirandola, Xiangyu Wang, Chao Zhou, Binjie
  Chu, Yijia Zhao, Bingjie Xu, Song Yu, and Hong Guo.
\newblock Long-distance continuous-variable quantum key distribution over
  202.81 km of fiber.
\newblock \emph{Phys. Rev. Lett.}, 125\penalty0 (1):\penalty0 010502, Jun 2020.
\newblock \doi{10.1103/PhysRevLett.125.010502}.

\bibitem[Mani et~al.(2021)Mani, Gehring, Grabenweger, \"Omer, Pacher, and
  Andersen]{Mani_21}
Hossein Mani, Tobias Gehring, Philipp Grabenweger, Bernhard \"Omer, Christoph
  Pacher, and Ulrik~Lund Andersen.
\newblock Multiedge-type low-density parity-check codes for continuous-variable
  quantum key distribution.
\newblock \emph{Phys. Rev. A}, 103:\penalty0 062419, 2021.
\newblock \doi{10.1103/PhysRevA.103.062419}.

\bibitem[Jiang et~al.(2017)Jiang, Huang, Huang, Lin, and Zeng]{Jiang_17}
Xue-Qin Jiang, Peng Huang, Duan Huang, Dakai Lin, and Guihua Zeng.
\newblock Secret information reconciliation based on punctured low-density
  parity-check codes for continuous-variable quantum key distribution.
\newblock \emph{Phys. Rev. A}, 95:\penalty0 022318, 2017.
\newblock \doi{10.1103/PhysRevA.95.022318}.

\bibitem[Wang et~al.(2017)Wang, Zhang, Li, Xu, Yu, and Guo]{Wang_17}
Xiangyu Wang, Yichen Zhang, Zhengyu Li, Bingjie Xu, Song Yu, and Hong Guo.
\newblock Efficient rate-adaptive reconciliation for continuous variable
  quantum key distribution.
\newblock \emph{Quantum Inform. Comput.}, 17\penalty0 (13\&14):\penalty0
  1123--1134, 2017.
\newblock \doi{10.26421/QIC17.13-14-4}.

\bibitem[Jeong et~al.(2022)Jeong, Jung, and Ha]{Jeong_22}
Suhwang Jeong, Hyunwoo Jung, and Jeongseok Ha.
\newblock Rate-compatible multi-edge type low-density parity-check code
  ensembles for continuous-variable quantum key distribution systems.
\newblock \emph{npj Quantum Inform.}, 8\penalty0 (6), 2022.
\newblock \doi{10.1038/s41534-021-00509-9}.

\bibitem[Milicevic et~al.(2018)Milicevic, Feng, Zhang, and Gulak]{Milicevic_18}
Mario Milicevic, Chen Feng, Lei~M. Zhang, and P.~Glenn Gulak.
\newblock Quasi-cyclic multi-edge ldpc codes for long-distance quantum
  cryptography.
\newblock \emph{npj Quantum Inform.}, 4\penalty0 (21), 2018.
\newblock \doi{10.1038/s41534-018-0070-6}.

\bibitem[Johnson et~al.(2016)Johnson, Chandrasetty, and Lance]{Johnson_16}
Sarah~J. Johnson, Vikram~A. Chandrasetty, and Andrew~M. Lance.
\newblock Repeat-accumulate codes for reconciliation in continuous variable
  quantum key distribution.
\newblock In \emph{2016 Australian Communications Theory Workshop (AusCTW)},
  pages 18--23, 2016.
\newblock \doi{10.1109/AusCTW.2016.7433603}.

\bibitem[Shirvanimoghaddam et~al.(2016)Shirvanimoghaddam, Johnson, and
  Lance]{Shirvanimoghaddam_16}
Mahyar Shirvanimoghaddam, Sarah~J. Johnson, and Andrew~M. Lance.
\newblock Design of raptor codes in the low snr regime with applications in
  quantum key distribution.
\newblock In \emph{2016 IEEE International Conference on Communications (ICC)},
  pages 1--6, 2016.
\newblock \doi{10.1109/ICC.2016.7510800}.

\bibitem[Zhou et~al.(2019)Zhou, Wang, Zhang, Zhang, Yu, and Guo]{Zhou_19}
Chao Zhou, Xiangyu Wang, Yichen Zhang, Zhiguo Zhang, Song Yu, and Hong Guo.
\newblock Continuous-variable quantum key distribution with rateless
  reconciliation protocol.
\newblock \emph{Phys. Rev. Appl.}, 12:\penalty0 054013, Nov 2019.
\newblock \doi{10.1103/PhysRevApplied.12.054013}.

\bibitem[Zhang et~al.(2004)Zhang, Wang, Son, and Kim]{Zhang_24}
Meixiang Zhang, Qiang Wang, Thara Son, and Sooyoung Kim.
\newblock Evaluation of adaptive reconciliation protocols for cv-qkd using
  systematic polar codes.
\newblock \emph{Quantum Inf. Process.}, 23\penalty0 (157), 2004.
\newblock \doi{10.1007/s11128-024-04371-4}.

\bibitem[Lucamarini et~al.(2018)Lucamarini, Yuan, Dynes, and
  Shields]{lucamarini2018overcoming}
Marco Lucamarini, Zhiliang~L Yuan, James~F Dynes, and Andrew~J Shields.
\newblock Overcoming the rate--distance limit of quantum key distribution
  without quantum repeaters.
\newblock \emph{Nature}, 557\penalty0 (7705):\penalty0 400--403, 2018.
\newblock \doi{10.1038/s41586-018-0066-6}.

\bibitem[Chen et~al.(2020)Chen, Zhang, Liu, Jiang, Zhang, Hu, Guan, Yu, Xu,
  Lin, Li, Chen, Li, You, Wang, Wang, Zhang, and Pan]{Chen_20}
Jiu-Peng Chen, Chi Zhang, Yang Liu, Cong Jiang, Weijun Zhang, Xiao-Long Hu,
  Jian-Yu Guan, Zong-Wen Yu, Hai Xu, Jin Lin, Ming-Jun Li, Hao Chen, Hao Li,
  Lixing You, Zhen Wang, Xiang-Bin Wang, Qiang Zhang, and Jian-Wei Pan.
\newblock Sending-or-not-sending with independent lasers: Secure twin-field
  quantum key distribution over 509 km.
\newblock \emph{Phys. Rev. Lett.}, 124:\penalty0 070501, Feb 2020.
\newblock \doi{10.1103/PhysRevLett.124.070501}.
\newblock URL \url{https://link.aps.org/doi/10.1103/PhysRevLett.124.070501}.

\bibitem[Wang et~al.(2022)Wang, Yin, He, Chen, Wang, Ye, Zhou, Fan-Yuan, Wang,
  Zhu, Morozov, Divochiy, Zhou, Guo, and Han]{Wang_22}
Shuang Wang, Zhen-Qiang Yin, De-Yong He, Wei Chen, Rui-Qiang Wang, Peng Ye, Yao
  Zhou, Guan-Jie Fan-Yuan, Fang-Xiang Wang, Yong-Gang Zhu, Pavel~V. Morozov,
  Alexander~V. Divochiy, Zheng Zhou, Guang-Can Guo, and Zheng-Fu Han.
\newblock Twin-field quantum key distribution over 830-km fibre.
\newblock \emph{Nature Photonics}, 16\penalty0 (2):\penalty0 154--161, Feb
  2022.
\newblock ISSN 1749-4893.
\newblock \doi{10.1038/s41566-021-00928-2}.

\bibitem[Liu et~al.(2023)Liu, Zhang, Jiang, Chen, Zhang, Pan, Ma, Dong, Xiong,
  Zhang, et~al.]{liu2023experimental}
Yang Liu, Wei-Jun Zhang, Cong Jiang, Jiu-Peng Chen, Chi Zhang, Wen-Xin Pan,
  Di~Ma, Hao Dong, Jia-Min Xiong, Cheng-Jun Zhang, et~al.
\newblock Experimental twin-field quantum key distribution over 1000 km fiber
  distance.
\newblock \emph{Phys. Rev. Lett.}, 130\penalty0 (21):\penalty0 210801, 2023.
\newblock \doi{10.1103/PhysRevLett.130.210801}.

\bibitem[Ferraz et~al.(2022)Ferraz, Subramaniyan, Chinthala, Andrade,
  Cavallaro, Nandy, Silva, Zhang, Purnaprajna, and Falcao]{Ferraz_22}
Oscar Ferraz, Srinivasan Subramaniyan, Ramesh Chinthala, João Andrade,
  Joseph~R. Cavallaro, Soumitra~K. Nandy, Vitor Silva, Xinmiao Zhang, Madhura
  Purnaprajna, and Gabriel Falcao.
\newblock A survey on high-throughput non-binary ldpc decoders: Asic, fpga, and
  gpu architectures.
\newblock \emph{IEEE Commun. Surv. Tutor.}, 24\penalty0 (1):\penalty0 524--556,
  2022.
\newblock \doi{10.1109/COMST.2021.3126127}.

\bibitem[Gallager(1963)]{Gallager_63}
Robert~G. Gallager.
\newblock \emph{Low-density parity-check codes}.
\newblock MIT Press, Cambridge, 1963.

\bibitem[MacKay and Neal(1996)]{MacKay_96}
David~J.C. MacKay and R.M. Neal.
\newblock Near shannon limit performance of low density parity check codes.
\newblock \emph{Electron. Lett.}, 32\penalty0 (18):\penalty0 1645--1646, 1996.
\newblock \doi{10.1049/el:19961141}.

\bibitem[MacKay(1999)]{MacKay_99}
David~J.C. MacKay.
\newblock Good error-correcting codes based on very sparse matrices.
\newblock \emph{IEEE Trans. Inf. Theory}, 45\penalty0 (2):\penalty0 399--431,
  1999.
\newblock \doi{10.1109/18.748992}.

\bibitem[Richardson et~al.(2001)Richardson, Shokrollahi, and
  Urbanke]{Richardson_01b}
Thomas~J. Richardson, Mohammad~Amin Shokrollahi, and R{\"u}diger~L. Urbanke.
\newblock Design of capacity-approaching irregular low-density parity-check
  codes.
\newblock \emph{IEEE Trans. Inf. Theory}, 47\penalty0 (2):\penalty0 619--637,
  2001.
\newblock \doi{10.1109/18.910578}.

\bibitem[Davey and MacKay(1998)]{Davey_98}
M.C. Davey and D.J.C. MacKay.
\newblock Low density parity check codes over {GF}(q).
\newblock In \emph{IEEE Information Theory Workshop (ITW)}, pages 70--71, 1998.
\newblock \doi{10.1109/ITW.1998.706440}.

\bibitem[Barnault and Declercq(2003)]{Barnault_03}
Lo\"{i}c Barnault and David Declercq.
\newblock Fast decoding algorithm for {LDPC} over {GF}($2^q$).
\newblock In \emph{ITW 2003, IEEE Inf. Theory Workshop}, pages 70--73. IEEE,
  2003.
\newblock \doi{10.1109/ITW.2003.1216697}.

\bibitem[Voicila et~al.(2010)Voicila, Declercq, Verdier, Fossorier, and
  Urard]{Voicila_10}
Adrian Voicila, David Declercq, Francois Verdier, Marc Fossorier, and Pascal
  Urard.
\newblock Low-complexity decoding for non-binary {LDPC} codes in high order
  fields.
\newblock \emph{IEEE Trans. Commun.}, 58\penalty0 (5):\penalty0 1365--1375,
  2010.
\newblock \doi{10.1109/TCOMM.2010.05.070096}.

\bibitem[Richardson and Urbanke(2001)]{Richardson_01a}
Thomas~J. Richardson and R{\"u}diger~L. Urbanke.
\newblock The capacity of low-density parity-check codes under message-passing
  decoding.
\newblock \emph{IEEE Trans. Inf. Theory}, 47\penalty0 (2):\penalty0 599--618,
  2001.
\newblock \doi{10.1109/18.910577}.

\bibitem[Chung et~al.(2001)Chung, Forney, Richardson, and Urbanke]{Chung_01}
Sae-Young Chung, Jr. Forney, G.D., Thomas~J. Richardson, and R{\"u}diger~L.
  Urbanke.
\newblock On the design of low-density parity-check codes within 0.0045 {dB} of
  the {Shannon} limit.
\newblock \emph{IEEE Commun. Lett.}, 5\penalty0 (2):\penalty0 58--60, 2001.
\newblock \doi{10.1109/4234.905935}.

\bibitem[Poulliat et~al.(2008)Poulliat, Fossorier, and Declercq]{Polliat_08}
Charly Poulliat, Marc Fossorier, and David Declercq.
\newblock Design of regular $(2,d_c)$-{LDPC} codes over {GF}($q$) using their
  binary images.
\newblock \emph{IEEE Trans. Commun.}, 56\penalty0 (10):\penalty0 1626--1635,
  2008.
\newblock \doi{10.1109/TCOMM.2008.060527}.

\bibitem[Kasai et~al.(2010)Kasai, Matsumoto, and Sakaniwa]{Kasai_10b}
Kenta Kasai, Ryutaroh Matsumoto, and Kohichi Sakaniwa.
\newblock Information reconciliation for {QKD} with rate-compatible non-binary
  {LDPC} codes.
\newblock In \emph{IEEE International Symposium On Information Theory and Its
  Applications (ISITA)}, pages 922--927, 2010.
\newblock \doi{10.1109/ISITA.2010.5649550}.

\bibitem[Mueller et~al.(2024)Mueller, Ribezzo, Zahidy, Oxenl{\o}we, Bacco, and
  Forchhammer]{Mueller_24}
Ronny Mueller, Domenico Ribezzo, Mujtaba Zahidy, Leif~Katsuo Oxenl{\o}we,
  Davide Bacco, and S{\o}ren Forchhammer.
\newblock Efficient information reconciliation for high-dimensional quantum key
  distribution.
\newblock \emph{Quantum Inf. Process.}, 23\penalty0 (5):\penalty0 195, May
  2024.
\newblock ISSN 1573-1332.
\newblock \doi{10.1007/s11128-024-04395-w}.

\bibitem[Kiktenko et~al.(2017)Kiktenko, Trushechkin, Lim, Kurochkin, and
  Fedorov]{Kiktenko_17}
E.~O. Kiktenko, A.~S. Trushechkin, C.~C.~W. Lim, Y.~V. Kurochkin, and A.~K.
  Fedorov.
\newblock Symmetric blind information reconciliation for quantum key
  distribution.
\newblock \emph{Phys. Rev. Appl.}, 8:\penalty0 044017, 2017.
\newblock \doi{10.1103/PhysRevApplied.8.044017}.

\bibitem[Liu et~al.(2020)Liu, Wu, and Huang]{Liu_20}
Zhihong Liu, Zhihao Wu, and Anqi Huang.
\newblock Blind information reconciliation with variable step sizes for quantum
  key distribution.
\newblock \emph{Sci. Rep.}, 10\penalty0 (1):\penalty0 171, 2020.
\newblock \doi{10.1038/s41598-019-56637-y}.

\bibitem[Andriyanova and Tillich(2012)]{Andriyanova_12}
Iryna Andriyanova and Jean-Pierre Tillich.
\newblock Designing a good low-rate sparse-graph code.
\newblock \emph{IEEE Trans. Commun.}, 60\penalty0 (11):\penalty0 3181--3190,
  2012.
\newblock \doi{10.1109/TCOMM.2012.082712.100205}.

\bibitem[Richardson and Urbanke(2004)]{Richardson_04}
Thomas~J. Richardson and R{\"u}diger~L. Urbanke.
\newblock Multi-edge type {LDPC} codes.
\newblock submitted IEEE IT, LTHC-REPORT-2004, 2004.

\bibitem[Johnson et~al.(2017)Johnson, Lance, Ong, Shirvanimoghaddam, Ralph, and
  Symul]{Johnson_17}
Sarah~J Johnson, Andrew~M Lance, Lawrence Ong, Mahyar Shirvanimoghaddam, T~C
  Ralph, and Thomas Symul.
\newblock On the problem of non-zero word error rates for fixed-rate error
  correction codes in continuous variable quantum key distribution.
\newblock \emph{New J. Phys.}, 19\penalty0 (2):\penalty0 023003, 2017.
\newblock \doi{10.1088/1367-2630/aa54d7}.

\bibitem[Jeong and Ha(2019)]{Suhwang_19}
Suhwang Jeong and Jeongseok Ha.
\newblock On the design of multi-edge type low-density parity-check codes.
\newblock \emph{IEEE Trans. Commun.}, 67\penalty0 (10):\penalty0 6652--6667,
  2019.
\newblock \doi{10.1109/TCOMM.2019.2927567}.

\bibitem[Pacher et~al.(2016)Pacher, Martinez-Mateo, Duhme, Gehring, and
  Furrer]{Pacher_16}
Christoph Pacher, Jesus Martinez-Mateo, Jörg Duhme, Tobias Gehring, and Fabian
  Furrer.
\newblock Information reconciliation for continuous-variable quantum key
  distribution using non-binary low-density parity-check codes, 2016.

\bibitem[Kasai et~al.(2011)Kasai, Declercq, Poulliat, and Sakaniwa]{Kasai_11}
Kenta Kasai, David Declercq, Charly Poulliat, and Kohichi Sakaniwa.
\newblock Multiplicatively repeated nonbinary {LDPC} codes.
\newblock \emph{IEEE Trans. Inf. Theory}, 57\penalty0 (10):\penalty0
  6788--6795, 2011.
\newblock \doi{10.1109/TIT.2011.2162259}.

\bibitem[Pearl(1988)]{Pearl_88}
Judea Pearl.
\newblock \emph{Probabilistic reasoning in intelligent systems: networks of
  plausible inference}.
\newblock Morgan Kaufmann, 1988.

\bibitem[Slepian and Wolf(1973)]{Slepian_73}
D.~Slepian and J.~Wolf.
\newblock Noiseless coding of correlated information sources.
\newblock \emph{IEEE Trans. Inf. Theory}, 19\penalty0 (4):\penalty0 471--480,
  1973.
\newblock \doi{10.1109/TIT.1973.1055037}.

\bibitem[Wyner(1975)]{Wyner_75}
A.~Wyner.
\newblock On source coding with side information at the decoder.
\newblock \emph{IEEE Trans. Inf. Theory}, 21\penalty0 (3):\penalty0 294--300,
  1975.
\newblock \doi{10.1109/TIT.1975.1055374}.

\bibitem[Liveris et~al.(2002)Liveris, Xiong, and Georghiades]{Liveris_02}
A.D. Liveris, Zixiang Xiong, and C.N. Georghiades.
\newblock Compression of binary sources with side information at the decoder
  using {LDPC} codes.
\newblock \emph{IEEE Commun. Lett.}, 6\penalty0 (10):\penalty0 440--442, 2002.

\bibitem[Declercq and Fossorier(2007)]{Declercq_07}
David Declercq and Marc Fossorier.
\newblock Decoding algorithms for nonbinary {LDPC} codes over {GF}($q$).
\newblock \emph{IEEE Trans. Commun.}, 55\penalty0 (4):\penalty0 633--643, 2007.
\newblock \doi{10.1109/TCOMM.2007.894088}.

\bibitem[Wei et~al.(2025)Wei, Garg, Nagai, Tomono, and Amano]{wei2025fpt}
Kaijie Wei, Devanshu Garg, Ryutaro Nagai, Takao Tomono, and Hideharu Amano.
\newblock Fpt-ems: An fpga implementation using nb-ldpc code for
  continuous-variable quantum key distribution.
\newblock In \emph{Proceedings of the 15th International Symposium on Highly
  Efficient Accelerators and Reconfigurable Technologies}, pages 117--125,
  2025.

\bibitem[Yang et~al.(2023)Yang, Yan, Yang, Lu, Lu, Cheng, Miao, and
  Li]{Yang_23}
Shenshen Yang, Zhilei Yan, Hongzhao Yang, Qing Lu, Zhenguo Lu, Liuyong Cheng,
  Xiangyang Miao, and Yongmin Li.
\newblock Information reconciliation of continuous-variables quantum key
  distribution: principles, implementations and applications.
\newblock \emph{EPJ Quantum Technol.}, 10\penalty0 (1):\penalty0 40, 2023.
\newblock \doi{10.1140/epjqt/s40507-023-00197-8}.

\bibitem[Laudenbach et~al.(2018)Laudenbach, Pacher, Fung, Poppe, Peev, Schrenk,
  Hentschel, Walther, and Hübel]{Laudenbach_18}
Fabian Laudenbach, Christoph Pacher, Chi-Hang~Fred Fung, Andreas Poppe,
  Momtchil Peev, Bernhard Schrenk, Michael Hentschel, Philip Walther, and
  Hannes Hübel.
\newblock Continuous-variable quantum key distribution with gaussian
  modulation—the theory of practical implementations (adv. quantum technol.
  1/2018).
\newblock \emph{Adv. Quantum Technol.}, 1\penalty0 (1):\penalty0 1870011, 2018.
\newblock \doi{10.1002/qute.201870011}.

\bibitem[Leverrier(2023)]{Leverrier_23}
Anthony Leverrier.
\newblock Information reconciliation for discretely-modulated
  continuous-variable quantum key distribution, 2023.
\newblock URL \url{https://arxiv.org/abs/2310.17548}.

\bibitem[Shokrollahi and Storn(2000)]{Shokrollahi_00}
A.~Shokrollahi and R.~Storn.
\newblock Design of efficient erasure codes with differential evolution.
\newblock In \emph{IEEE International Symposium on Information Theory (ISIT)},
  pages 1--5, 2000.
\newblock \doi{10.1109/ISIT.2000.866295}.

\bibitem[Rathi and Urbanke(2005)]{Rathi_05}
V.~Rathi and R.~Urbanke.
\newblock Density evolution, thresholds and the stability condition for
  non-binary ldpc codes.
\newblock \emph{IEE Proceedings - Communications}, 152:\penalty0 1069--1074(5),
  2005.
\newblock \doi{10.1049/ip-com:20050230}.

\bibitem[Hu et~al.(2005)Hu, Eleftheriou, and Arnold]{Hu_05}
X.-Y. Hu, E.~Eleftheriou, and D.~M. Arnold.
\newblock Regular and irregular progressive edge-growth {Tanner} graphs.
\newblock \emph{IEEE Trans. Inf. Theory}, 51\penalty0 (1):\penalty0 386--398,
  2005.
\newblock \doi{10.1109/TIT.2004.839541}.

\bibitem[Tomamichel et~al.(2017)Tomamichel, Martinez-Mateo, Pacher, and
  Elkouss]{Tomamichel_17}
Marco Tomamichel, Jesus Martinez-Mateo, Christoph Pacher, and David Elkouss.
\newblock Fundamental finite key limits for one-way information reconciliation
  in quantum key distribution.
\newblock \emph{Quantum Inf. Process.}, 16\penalty0 (11):\penalty0 280, 2017.
\newblock \doi{10.1007/s11128-017-1709-5}.

\bibitem[Lodewyck et~al.(2007)Lodewyck, Bloch, Garc\'{\i}a-Patr\'on, Fossier,
  Karpov, Diamanti, Debuisschert, Cerf, Tualle-Brouri, McLaughlin, and
  Grangier]{Lodewyck_07}
J\'er\^ome Lodewyck, Matthieu Bloch, Ra\'ul Garc\'{\i}a-Patr\'on, Simon
  Fossier, Evgueni Karpov, Eleni Diamanti, Thierry Debuisschert, Nicolas~J.
  Cerf, Rosa Tualle-Brouri, Steven~W. McLaughlin, and Philippe Grangier.
\newblock Quantum key distribution over
  $25\phantom{\rule{0.3em}{0ex}}\mathrm{km}$ with an all-fiber
  continuous-variable system.
\newblock \emph{Phys. Rev. A}, 76:\penalty0 042305, Oct 2007.
\newblock \doi{10.1103/PhysRevA.76.042305}.

\bibitem[Leverrier et~al.(2010)Leverrier, Grosshans, and
  Grangier]{Leverrier_10}
Anthony Leverrier, Fr\'ed\'eric Grosshans, and Philippe Grangier.
\newblock Finite-size analysis of a continuous-variable quantum key
  distribution.
\newblock \emph{Phys. Rev. A}, 81:\penalty0 062343, Jun 2010.
\newblock \doi{10.1103/PhysRevA.81.062343}.

\end{thebibliography}

\end{document}